\documentclass{aa}  
\usepackage{multirow}
\usepackage{graphicx}
\usepackage{txfonts}
\usepackage{lipsum}
\usepackage{subcaption}
\usepackage{lscape}           
\usepackage{placeins}          
\usepackage[nohyperlinks]{acronym} 
\usepackage{color}
\usepackage{xcolor}
\usepackage{orcidlink}
\usepackage{comment}

\newcommand{\sherry}[1]{\textcolor{orange}{\textbf{[Sherry: #1]}}}

\newcommand{\todo}[1]{\textcolor{red}{\textbf{[To Do: #1]}}}

\newacro{mge}[MGE]{multi-Gaussian expansion}
\newacro{psf}[PSF]{point spread function}
\newacro{sie}[SIE]{singular isothermal ellipsoid}
\newacro{sis}[SIS]{singular isothermal sphere}

\makeatletter
\let\linenumbers\relax
\makeatother

\begin{document}

   \title{HOLISMOKES XX. Lens models of binary lens galaxies with five images of Supernova Winny}

\author{L.~R.~Ecker\inst{\ref{usm},\ref{mpe}}\thanks{L.Ecker@campus.lmu.de}\orcidlink{0009-0005-3508-2469}
\and
A.~G.~Schweinfurth\inst{\ref{tum},\ref{mpa}}\thanks{allansch@mpa-garching.mpg.de}\orcidlink{0000-0002-8274-7196}
\and
R.~Saglia\inst{\ref{mpe},\ref{usm}}\orcidlink{0000-0003-0378-7032}
\and 
L.~Deng\inst{\ref{tum},\ref{mpa}}
\orcidlink{0009-0009-9255-920X}
\and
S.~H.~Suyu\inst{\ref{tum},\ref{mpa}}
\orcidlink{0000-0001-5568-6052} 
\and
C.~Saulder\inst{\ref{mpe},\ref{usm}}\orcidlink{0000-0002-0408-5633}
\and
J.~Snigula\inst{\ref{mpe}}
\and
R.~Bender\inst{\ref{usm},\ref{mpe}}\orcidlink{0000-0001-7179-0626}
\and
R.~Ca\~nameras\inst{\ref{aix}}
\and
T.-W.~Chen\inst{\ref{janet}}\orcidlink{0000-0002-1066-6098}
\and
A.~Galan\inst{\ref{swiss}}\orcidlink{0000-0003-2547-9815}
\and
A.~Halkola\inst{\ref{tuusula}}
\and
E.~Mamuzic\inst{\ref{tum}, \ref{mpa}}
\and
A.~Melo\inst{\ref{eso}}
\and
S.~Schuldt\inst{\ref{stefan1},\ref{stefan2},\ref{stefan3}}\orcidlink{0000-0003-2497-6334}
\and
S.~Taubenberger\inst{\ref{tum},\ref{mpa}}
}

\institute{
    University Observatory Munich, Faculty of Physics, Ludwig-Maximilians-Universit\"at, Scheinerstr. 1, 81679 Munich, Germany
    \href{mailto:L.Ecker@campus.lmu.de}{\tt L.Ecker@campus.lmu.de} \label{usm}
    \and
    Max Planck Institute for Extraterrestrial Physics, Giessenbachstr. 1, 85748 Garching, Germany \label{mpe}
    \and
    Technical University of Munich, TUM School of Natural Sciences, Physics Department,  James-Franck-Stra{\ss}e 1, 85748 Garching, Germany \href{mailto:allansch@mpa-garching.mpg.de}{\tt allansch@mpa-garching.mpg.de}  \label{tum}
    \and
    Max Planck Institute for Astrophysics, Karl-Schwarzschild-Stra{\ss}e 1, 85748 Garching, Germany \label{mpa}
    \and   
    Aix-Marseille Université, CNRS, CNES, LAM, Marseille, France \label{aix}
    \and
    Graduate Institute of Astronomy, National Central University,
    300 Jhongda Road, 32001 Jhongli, Taiwan\label{janet}
    \and
    Department of Astronomy, University of Geneva, ch. d'Ecogia 16, 1290 Versoix, Switzerland\label{swiss}
    \and
    Py\"orrekuja 5 A, 04300 Tuusula, Finland \label{tuusula}
    \and
    European Southern Observatory, Karl-Schwarzschild-Strasse 2, D-85748 Garching bei München, Germany\label{eso}
    \and
    Finnish Centre for Astronomy with ESO (FINCA), University of Turku, FI-20014 Turku, Finland\label{stefan1}
    \and
    Department of Physics, P.O. Box 64, University of Helsinki, FI-00014 Helsinki, Finland\label{stefan2}
    \and
    INAF - IASF Milano, via A. Corti 12, I-20133 Milano, Italy\label{stefan3}
}

   \date{Received Xxxxxxx xx, 2026}

\abstract{Strongly lensed supernovae (SNe) provide a powerful way to study cosmology, SNe and galaxies. Modelling the lens system is key to extracting astrophysical and cosmological information. We present adaptive-optics-assisted high-resolution images of the recently discovered SN Winny (SN 2025wny) in the $J$ and $K$ filters obtained with the Large Binocular Telescope. The high-resolution adaptive optics LBT imaging confirms the presence of a fifth point source, whose colour is consistent with that of the other SN images at similar phases, while lens modelling robustly supports its interpretation as an additional image of SN~Winny. We measure the positions of the five SN images with uncertainties varying between 1 and 14 milliarcseconds. Using the five SN image positions as constraints and the centroids of the lens light distributions as priors for their mass centroids, we build the first mass models using two different pieces of software, \texttt{lenstronomy} and \texttt{GLEE}.  We explored three classes of mass models for the two lens galaxies G1 and G2 involving singular isothermal sphere (SIS), singular isothermal ellipsoid (SIE) and external shear profiles. The optimal model class, based on the Bayesian Information Criterion, is an SIE for G1, an SIS for G2, and an external shear for both \texttt{lenstronomy} and \texttt{GLEE}. From the lens modelling, we infer the enclosed masses within the Einstein radius as $M_{\rm G1}(<\theta_{\rm E}) = 4.61^{+0.06}_{-0.04} \times 10^{11}\,M_\odot$ for G1 and $M_{\rm G2}(<\theta_{\rm E}) = 1.01 \pm 0.02 \times 10^{11}\,M_\odot$ for G2. The lensing configuration by the two lens galaxies can produce two additional magnified SN images beyond the five observed ones; the exclusion of such model configurations further constrains the lens mass model parameters. Our model fits to the observed image positions with an RMS of $\sim 0.0012\, \arcsec - 0.0025\, \arcsec$, within the observed positional uncertainties and without additional predicted SN images.
  The predicted magnifications of the multiple images vary between $\sim1.6$ (for the faintest fifth image E) to $\sim10$ (for the brightest image A). The predicted relative lensing magnifications of the multiple images do not match that of the observed within $2\sigma$ uncertainties. The differences in the relative magnifications could be due to millilensing and microlensing effects. Our mass models form the basis for future analyses of this unique system.
}


   \maketitle

\section{Introduction}
\label{sec:intro}

Strong gravitational lensing of transients and time-varying sources, such as supernovae (SNe) and quasars, provides a powerful probe of the mass distribution of the lens galaxy and a direct method for measuring the Hubble constant, $H_0$, via time-delay cosmography \citep{Refsdal1964}. While lensed quasars have been the primary targets for such studies over the past decades \citep[e.g.,][]{Wong2020, TDCosmoMilestone}, strongly lensed supernovae offer distinct advantages.  For example, their finite duration allows for a robust measurement of their time delay in a short timescale compared to quasars. Despite their utility, lensed supernovae remain rare phenomena, though the discovery and characterisation of these systems have accelerated recently due to high-cadence surveys \citep[see reviews by, e.g.,][]{Oguri2019, Suyu2024}. 

\begin{figure}[tbp!]
    \centering
    \includegraphics[width=\linewidth]{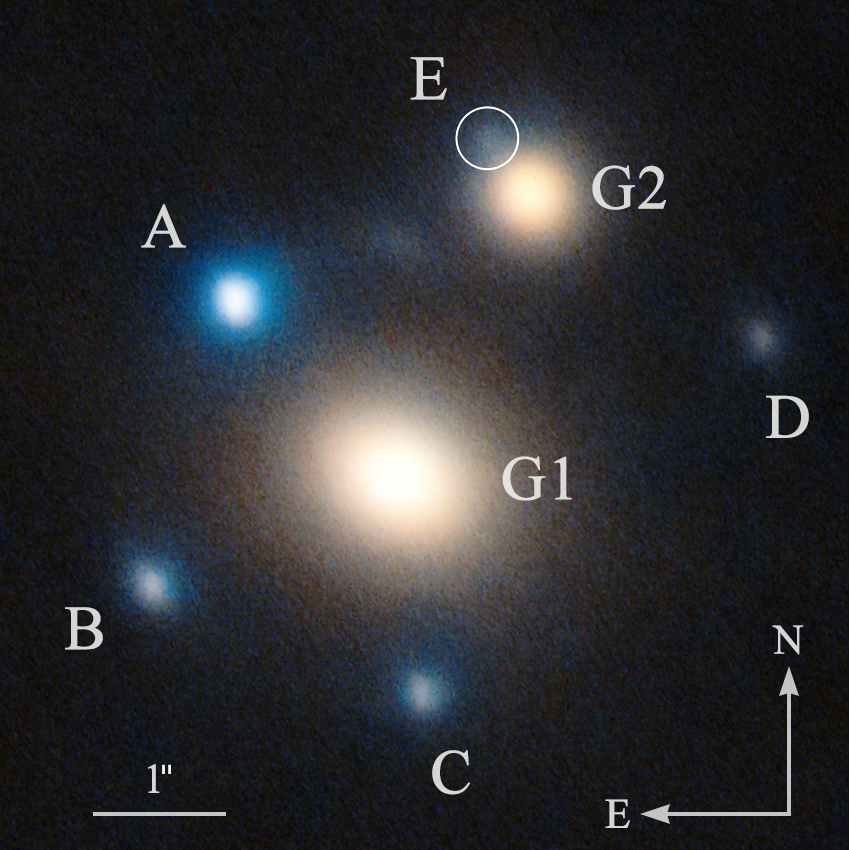}
    \caption{Colour composite image of the lens system of SN Winny, constructed using $J$ and $K$ band data from the Large Binocular Telescope (LBT). The primary lensing galaxies are labelled G1 and G2. The multiple images of the supernova are marked A--D, while E denotes the location of the possible fifth image reported by \citet{Aryan+2025}.}
    \label{fig:colourcomposite}
\end{figure}

One exciting discovery is SN~Winny, the first strongly lensed superluminous supernova at $z=2$ 
\citep{Taubenberger2025, Johansson2025}. 
This is the third galaxy-scale lensed supernova with spatially-resolved multiple SN images after iPTF16geu \citep[e.g.,][]{Goobar+2017, More+2017, Dhawan+2020, Baltasar+2026} and SN Zwicky \citep[e.g.,][]{Goobar+2023, Pierel+2023, Larison+2025}. SN~Winny is a remarkable system featuring a binary lens configuration composed of two lens galaxies. Initial observations confirmed the presence of four images \citep{Perley_astronote2025}, and Canada-France-Hawaii Telescope (CFHT) follow-up suggested a possible fifth image \citep{Aryan+2025}. The system offers a unique opportunity to study the mass distribution of a binary lens using the precise astrometric constraints from multiple SN images.

In this work, we present a detailed analysis of SN~Winny. We utilise high-resolution adaptive optics (AO) imaging obtained with the Large Binocular Telescope (LBT) in the near-infrared $J$ and $K$ bands. The high spatial resolution of this data set enables precise astrometric measurements of the lensed SN images and robust lens-light subtraction. In addition, we utilise these data to verify the fifth image candidate (Image E) through photometric consistency (colour analysis) and lens mass modelling.

We perform lens mass modelling using two independent software packages, the Gravitational Lens Efficient Explorer \citep[\texttt{GLEE};][]{Suyu2010, Suyu2012} and \texttt{lenstronomy} \citep{Birrer2018}, to derive the properties of the deflectors. We explore different combinations of mass parameterisations, including singular isothermal sphere (SIS) and singular isothermal ellipsoid (SIE) profiles, to identify the model that best reproduces the observed astrometry and image multiplicity. By combining these results, we obtain a mass model for this system that fits our astrometric constraints. While not yet of cosmography grade, since that requires measuring the lens radial profile slope  (instead of the isothermal assumption) which needs more observational constraints, this model lays the groundwork for future cosmographic analysis.

The paper is organised as follows. In Sect.~\ref{sec:obs}, we describe the LBT observations: data reduction, the lens light subtraction methodology, point spread function (PSF) reconstruction, the derived astrometric and photometric measurements and a brief discussion on the colour of the putative fifth image. In Sect.~\ref{sec:mass_model}, we detail the mass modelling strategy, including the parameterisation of the lens galaxies and the priors used. We describe the procedure to model this system and present the results, followed by a discussion in Sect.~\ref{sec:discussion}. The parameter values are reported as the median, with uncertainties given by the 16th and 84th percentiles, unless otherwise stated. We assume a flat $\Lambda$CDM cosmology with $\Omega_{\rm m} = 0.3 = 1- \Omega_\Lambda$ and $H_0 = 70\,\mathrm{km\,s^{-1}\,Mpc^{-1}}$, chosen to remain agnostic regarding the Hubble tension.

\section{Observations}\label{sec:obs}
\subsection{Data reduction}
We observed the system on the 25th and 26th of November 2025, using the adaptive optics system of the LBT with a natural guide star and the LUCI instruments in imaging mode with the N30 camera delivering 0.015\arcsec\ pixels and a field of view of 30\arcsec\,$\times$\,30\arcsec. The scripts controlling the telescope and LUCI operations were prepared using the LBTO OT software\footnote{\url{https://scienceops.lbto.org/script-preparation/ot-installation/}}. We collected a series of 60 sec $J$ and $K$ dithered exposures in good atmospheric and natural seeing conditions. Out of this collection, we selected a
sample of 29 $J$- and 30 $K$-band images observed with LUCI1 mounted on the left mirror of the LBT that delivered the sharpest PSF, with on average a full width at half maximum (FWHM) of 0.30\arcsec\ and 0.22\arcsec\ in the $J$ and $K$ band, respectively.
The images were reduced using MIDAS\footnote{\url{https://www.eso.org/sci/software/esomidas}}. After flatfielding, we median combined the selected images, erasing the sources thanks to the dithering and therefore delivering the sky background. After subtracting it from each frame, we aligned the images and the squares of their values, respectively, to produce the final total images and their errors, achieving signal-to-noise ratios per source pixel between five and thirty. Figure \ref{fig:colourcomposite} shows the colour image from these observations, with the SN and the lens galaxies labelled.

\subsection{Lens light subtraction}\label{sec: lenslightsub}
Accurate subtraction of the foreground lens galaxy light is essential to determine the positions of the SN images precisely, as the underlying lens light must be removed carefully to avoid biasing the measured positions of multiple images. Multi-Gaussian expansion (MGE) light profiles were used, which are capable of modelling the light distributions of galaxies \citep{cappellari02}, and was shown to perform well for strong lenses \citep{he24}.

The lens system contains two galaxies, hereafter referred to as G1 and G2. Visual inspection of the imaging data shows that the isophotes of the two galaxies do not significantly overlap and are overall smooth and unperturbed. This allows the light distributions of G1 and G2 to be modelled independently.

To model the light of G1, we mask the light from G2, the supernova images, and the lensed arc from the SN host galaxy. The remaining galaxy light was fitted using two independent MGE components, each consisting of 30 Gaussian profiles. Galaxy G2 was treated analogously, with G1, the supernova images, and the lensed arc masked before modelling its light distribution with two Gaussian sets of identical complexity.

In the MGE formalism, the surface brightness distribution of a galaxy is expressed as a sum of two-dimensional Gaussian components,
\begin{equation}
I(x, y) = \sum_{i=1}^{N} G_i(x, y),
\end{equation}
where $G_i(x, y)$ denotes the $i$-th Gaussian profile. Each Gaussian is given by
\begin{equation}
G_i(x, y) = I_i \exp\left(-\frac{R_i^2(x, y)}{2\sigma_i^2}\right),
\end{equation}
with $I_i$ and $\sigma_i$ representing the central intensity and width of the $i$-th Gaussian component, respectively. The quantity $R_i(x, y)$ is the elliptical radius, defined as
\begin{equation}
R_i(x, y) = \sqrt{x'^2 + \left(\frac{y'}{q_i}\right)^2},
\end{equation}
where $q_i$ is the axis ratio of the Gaussian. The rotated coordinates $(x', y')$ are given by
\begin{align}
x' &= \cos\phi_i \, (x - x_{\mathrm{c},i}) + \sin\phi_i \, (y - y_{\mathrm{c},i}), \\
y' &= \cos\phi_i \, (y - y_{\mathrm{c},i}) - \sin\phi_i \, (x - x_{\mathrm{c},i}),
\end{align}
with $(x_{\mathrm{c},i}, y_{\mathrm{c},i})$ denoting the Gaussian centre and $\phi_i$ the position angle measured east of north.

Within each Gaussian set, all components share a common axis ratio, position angle, and centroid. The Gaussian widths $\sigma$ were fixed and distributed logarithmically, spanning the range \[\sigma \in [0.001\, \arcsec, 5\, \arcsec].\] This logarithmic spacing allows the model to capture both the compact central light distribution and the extended low-surface-brightness emission.

Initial estimates for the galaxy centres were obtained using the \texttt{photutils} \citep{photutils} package, employing the \texttt{find\_center} method from the \texttt{EllipseGeometry} class. During the modelling, the central positions were assigned uniform priors within a square region of side length $0.2\, \arcsec$ centred on these initial estimates. The galaxy ellipticities were parameterised in terms of the components
\begin{equation}
\epsilon_1 = \frac{1 - q}{1 + q} \sin(2\phi), \qquad
\epsilon_2 = \frac{1 - q}{1 + q} \cos(2\phi), \label{eq: cart_ellip}
\end{equation}
where $q$ is the axis ratio and $\phi$ the position angle. Uniform priors were adopted for both ellipticity components in the range $[-0.7, 0.7]$.

The lens light modelling was performed using \texttt{PyAutoLens} \citep{Nightingale2021}, which optimises the amplitudes $I_i$ of the Gaussian components while keeping their widths and shared geometric parameters fixed. The best-fitting MGE models for G1 and G2 were combined and subtracted from the data.

\begin{figure}[htbp!]
\centering
\includegraphics[angle=0,width=1.0\hsize]{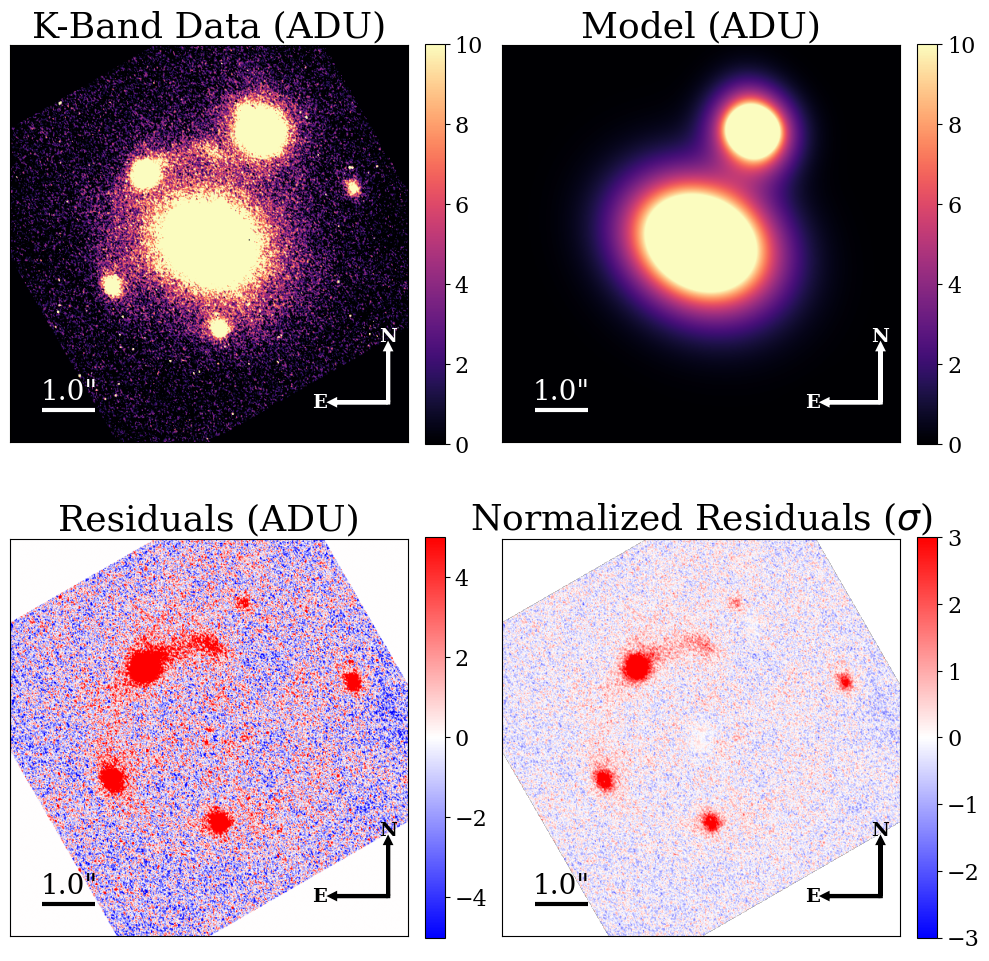}
\caption{Lens light subtraction using \texttt{PyAutoLens}. \textbf{Top left:} K-band imaging data. \textbf{Top right:} Combined multi-Gaussian expansion (MGE) models for the two lens galaxies G1 and G2. \textbf{Bottom left:} Residuals after subtracting the MGE models from the data. \textbf{Bottom right:} Normalised residuals in units of $\sigma$, showing no significant coherent structure (besides the SN images and the arc from the host galaxy) indicating that the lens light has been successfully removed. }\label{fig:lens_subtraction}

\end{figure}

Figure~\ref{fig:lens_subtraction} shows the results of the lens light subtraction. The top row displays the original $K$-band data and the combined MGE model, while the bottom row presents the absolute and normalised residuals. The residual images show no significant large-scale coherent structures, and the normalised residuals are largely consistent with noise, indicating that the lens light has been successfully removed without introducing systematic artefacts that could bias subsequent lens modelling.

\subsection{PSF reconstruction}

The PSF was reconstructed for each photometric band using the \texttt{STARRED} package \citep{Michalewicz2023, Millon_2024}, which models the PSF through a regularised, multi-scale approach. Due to the lack of stars in the field of view, the reconstruction was derived directly from $0.75\arcsec\times0.75\arcsec$ cutouts of the four brightest supernova images (A–D). This was performed only after the preliminary subtraction of the lens galaxy light discussed in Sect.~\ref{sec: lenslightsub}.

\begin{figure}[tbp!]
    \centering
    \includegraphics[width=\linewidth]{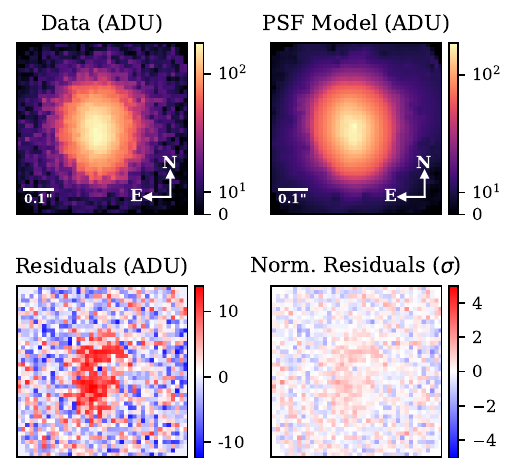}
    \caption{$K$-band PSF reconstruction for SN Winny image A. The panels display the observed data (top left), the PSF model fit to the data (top right), the absolute residuals (bottom left), and the normalised residuals (bottom right).}
    \label{fig:PSF_fit}
\end{figure}

We adopted a PSF model consisting of two components: an analytical Moffat profile to capture the seeing-limited core, and a pixelated grid to model finer, non-analytic asymmetries. We employed a supersampling factor of three relative to the native pixel scale of $0.015\, \arcsec$ and assumed spatial invariance across the field of view over the lensed images. The reconstruction proceeded in two steps: first, we fitted the Moffat profile to the supernova cutouts to obtain initial parameter estimates; second, we iteratively refined the pixel-based component using regularisation in wavelet space to promote smoothness while preserving structural details. An example of the PSF model and a fit to the $K$-band data of SN Winny image A is shown in Fig. \ref{fig:PSF_fit}.

\subsection{Astrometric and photometric measurements}\label{sec: astro+photo}

We measured the astrometric positions and photometry of the five supernova images (including image E) independently in the $K$ band and the $J$ band by fitting a modelled PSF to the surface brightness at the location of SN Winny. For this, we used \texttt{GLEE} and employed a combination of simulated annealing minimisation \citep{Kirkpatrick1983} and Markov chain Monte Carlo sampling \citep[MCMC;][]{Dunkeley2005} to model the surface brightness distribution. We explored the posterior distributions of the centroids and amplitude using the lens-light-subtracted data discussed in Sect.~\ref{sec: lenslightsub}.

Systematic uncertainties arising from PSF reconstruction choices in \texttt{STARRED} were estimated using a set of eight PSF models generated for each band. We varied the analytical component between circular and elliptical Moffat profiles and applied distinct regularisation schemes for each geometry. Specifically, we varied the regularisation strength of the highest frequency wavelets, $\lambda_{\rm hf}$, and the strength for all other scales, $\lambda_{\rm scales}$, within the grid $(\lambda_{\rm hf}, \lambda_{\rm scales}) \in \{3, 5\} \times \{3, 5\}$.

The final parameter values for each band were derived by marginalising over the eight chains, subject to an outlier rejection step. We discarded chains deviating from the median by more than $5\sigma$, calculated via the median absolute deviation, on a per-SN-image basis. We observed that the statistical uncertainty from MCMC sampling is comparable to the systematic scatter introduced by the different PSF models. We combined the posterior samples from all remaining configurations to incorporate both variance sources into the final error budget. A Gaussian fit was applied to this combined distribution, and the reported spatial uncertainty was symmetrised using the geometric mean of the fitted widths, $\sigma=\sqrt{\sigma_x\sigma_y}$.

Finally, we aligned the astrometric positions in the $J$ band using the $K$ band as reference frame by computing the rigid transformation (rotation and translation) that minimised the weighted squared residuals, $\chi_{\rm align}^2$, of the brightest four image positions (A--D). The weights were defined as the inverse combined variance, $w_i = (\sigma_{{\rm J},i}^2 + \sigma_{{\rm K},i}^2)^{-1}$ for $i = $\{A, B, C, D\}, to ensure the alignment was driven by the most precisely measured positions in both bands. The resulting astrometry and PSF amplitudes of the SN images in the $K$ and $J$ bands are listed in Table~\ref{tab: astrometry}.

\begin{table*}
\caption{Observed light centroids of the lens galaxies (G1, G2), relative astrometry of the multiple images of SN~Winny (A--E), and the total fluxes in the $J$ and $K$ bands.}
\label{tab: astrometry}
\centering
\begin{tabular}{l c c c c c c}
\hline\hline
\noalign{\smallskip}
 & \multicolumn{3}{c}{$J$ band} & \multicolumn{3}{c}{$K$ band} \\
Object & $\theta_1$ ($\arcsec$) & $\theta_2$ ($\arcsec$) & \shortstack{ Total flux (ADU)} & $\theta_1$ ($\arcsec$) & $\theta_2$ ($\arcsec$) & \shortstack{Total flux (ADU)} \\
\noalign{\smallskip}
\hline
\noalign{\smallskip}
G1 & $\phantom{-}0.0241 \pm 0.0025$ & $\phantom{-}0.0005\pm 0.0032$ & $333221\pm 577$ & $\phantom{-}0.0000 \pm 0.0031$ & $\phantom{-}0.0000 \pm 0.0012$ & $680020 \pm 825$ \\
G2 & $\phantom{-}0.9913 \pm 0.0017$ & $\phantom{-}2.1292 \pm 0.0016$ & $83538 \pm 289$ & $\phantom{-}0.9793 \pm 0.0023$ & $\phantom{-}2.1138 \pm 0.0010$ & $178424 \pm 927$ \\
A & $-1.1898 \pm 0.0024$ & $\phantom{-}1.3233 \pm 0.0024$ & $34657 \pm 248$ & $-1.1876 \pm 0.0012$ & $\phantom{-}1.3204 \pm 0.0012$ & $49677 \pm 332$ \\
B & $-1.8046 \pm 0.0036$ & $-0.8061 \pm 0.0036$ & $9941 \pm 147$ & $-1.8051 \pm 0.0024$ & $-0.8020 \pm 0.0024$ & $19336 \pm 297$ \\
C & $\phantom{-}0.2174 \pm 0.0096$ & $-1.6169 \pm 0.0096$ & $7571 \pm 521$ & $\phantom{-}0.2097 \pm 0.0036$ & $-1.6057 \pm 0.0036$ & $15221 \pm 339$ \\
D & $\phantom{-}2.7494 \pm 0.0060$ & $\phantom{-}1.0415 \pm 0.0060$ & $4109 \pm 102$ & $\phantom{-}2.7381 \pm 0.0060$ & $\phantom{-}1.0459 \pm 0.0060$ & $8206 \pm 232$ \\
E & $\phantom{-}0.6946 \pm 0.0120$ & $\phantom{-}2.5288 \pm 0.0120$ & $1855 \pm 100$ & $\phantom{-}0.6791 \pm 0.0144$ & $\phantom{-}2.5362 \pm 0.0144$ & $3085 \pm 214$ \\
\noalign{\smallskip}
\hline
\end{tabular}
\tablefoot{Positions are measured relative to the mean centre of light of G1 in the K band, with $\theta_1$ oriented West and $\theta_2$ North.}
\end{table*}

Since systematic residuals remained following this rotation, the pure statistical uncertainties were effectively underestimated. To derive realistic parameter errors, we scaled the uncertainties by a constant factor ($k=1.2$) such that the reduced $\chi_{\rm align}^2$ of the rotation fit was approximately unity.

\subsection{Colour of the SN images}
\begin{figure}[tbp!]
     \centering
     \includegraphics[width=0.8\linewidth]{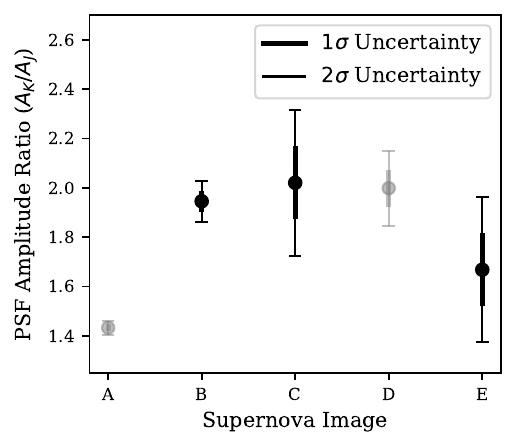}
    \caption{PSF amplitude ratios ($A_K / A_J$, colour ratios) of the five lensed images of SN Winny. The data points represent the mean ratios derived from the posterior distributions of the PSF amplitude. Thick and thin error bars correspond to the $1\sigma$ and $2\sigma$ uncertainties, respectively. Images A and D (shown in grey) are excluded from the consistency check due to the source in image A not being compact and possible time-delay induced colour evolution, respectively. The candidate image E agrees with the colours of images B and C within the $2\sigma$ confidence interval, supporting its identification as a fifth image of SN Winny.}
    \label{fig: colour}
 \end{figure}

We additionally analysed the relative colours ($K/J$ flux ratios) across the lensed images to verify that candidate image E has a colour consistent with the additional lensed images. We combined the PSF amplitude constraints from the $K$ band and the $J$ band by drawing $10,000$ random samples from their distributions to compute the probabilistic flux ratios and associated uncertainties, shown in Fig.~\ref{fig: colour}. For the candidate image E to be a SN image, its $K/J$ flux ratio value is expected to be consistent with that of images B and C. We exclude image D from this comparison because, given the ``cusp-like" SN image configuration, it is the first arrival image and could be significantly more evolved; consequently, its observed phase (and hence colour) is not necessarily expected to match that of the remaining images. We also exclude image A because the host galaxy does not appear compact in this image. In contrast, the host appears compact and contained within the PSF in images B, C, and E. While the measured photometry in these images is a superposition of both the SN and the host galaxy, the observed colour is expected to remain consistent. This is because the compact lensed morphology ensures that the host light is not sheared away from the supernova position; consequently, both sources likely experience similar lensing magnifications, thus preserving the intrinsic flux ratio between bands.

We found that the colour of candidate image E agrees with that of images B and C within $2\sigma$. This agreement serves as an important consistency check, supporting the identification of image E. Furthermore, \citet{Aryan+2025} detected a transient consistent with the position of image E, providing strong evidence that this is indeed a lensed SN image created by the presence of the secondary lens galaxy G2 in addition to the primary lens G1. Preliminary lens models using the archival Canada-France-Hawaii-Telescope images shown in, e.g., \citet{Taubenberger2025} predicted a fifth image in that location as well. Taken together, this mounting evidence suggests that candidate E is highly likely a lensed SN image. We therefore proceed to model the lens mass distribution with the five SN image positions as constraints.

\section{Lens Mass Model}\label{sec:mass_model}

\subsection{Mass parameterisation}\label{sec: massparam}

The lens mass distribution was modelled using two main deflectors, corresponding to the primary lens galaxy (G1) and its companion (G2). We assumed that both galaxies lie in the same lens plane, as they have consistent spectroscopic redshifts of $z=0.375$ \citep{desicollaboration2025datarelease1dark, Taubenberger2025}. 

To describe the mass profiles, we adopted a singular isothermal ellipsoid (SIE), corresponding to a pseudo-isothermal elliptical mass distribution (PIEMD; \citealt{Kassiola1993}) with no core. The dimensionless surface mass density (convergence), $\kappa$, at a position $(\theta_1, \theta_2)$ relative to the profile centre is given by:
\begin{equation}
    \kappa_\text{SIE} = \frac{\theta_{\rm E}}{2r_{\rm em}},
    \label{eq:piemd_convergence}
\end{equation}

\noindent where $\theta_{\rm E}$ represents the lens strength parameter, also called Einstein radius. The variable $r_{\text{em}}$ represents the elliptical radius, which serves as the radial coordinate in the elliptical frame. It is defined in a coordinate system aligned with the major and minor axes as:

\begin{equation}
    r_{\text{em}}^2 = \frac{\theta_1^2}{(1+e)^2} + \frac{\theta_2^2}{(1-e)^2},
\end{equation}

\noindent where $e$ is the ellipticity, which is related to the minor-to-major axis ratio $q$ and the ellipticity cartesian ellipticity components $\epsilon_1$ and $\epsilon_2$, defined as in eq. \eqref{eq: cart_ellip}, by 
\begin{equation}
    e = \frac{1-q}{1+q} = \sqrt{\epsilon_1^2+\epsilon_2^2}.
\end{equation}

This parameterisation allows us to recover the standard singular models used in strong lensing. With non-zero ellipticity ($e \neq 0$), the profile becomes a singular isothermal ellipsoid (SIE). If we further restrict the geometry to spherical symmetry ($e=0$, corresponding to $q=1$), the profile simplifies to the singular isothermal sphere (SIS).  

In addition to the lens galaxies, we include a constant external shear which, following \cite{Suyu2013}, can be parametrised by the following potential form:
\begin{equation}
    \psi_{\rm ext} (\theta, \phi) = \frac{1}{2} \gamma_{\rm ext}\theta^2 \cos 2(\phi - \phi_{\rm ext}),
\end{equation}
where ($\theta$, $\phi$) are the polar coordinates, $\gamma_{\rm ext}$ denotes the shear strength, and $\phi_{\rm ext}$ represents the shear position angle. An angle of $\phi_{\rm ext} = 0^{\circ}$ indicates a shear aligned with the $\theta_1$ axis and an angle of $\phi_{\rm ext} = 90^{\circ}$ indicates a shear aligned with the $\theta_2$ axis. Equivalently, this potential can be expressed in terms of the Cartesian shear components $\gamma_1 = \gamma_{\rm ext} \cos 2\phi_{\rm ext}$ and $\gamma_2 = \gamma_{\rm ext} \sin 2\phi_{\rm ext}$ as:
\begin{equation}
    \psi_{\rm ext}(\theta_1, \theta_2) = \frac{1}{2} \gamma_1 (\theta_1^2 - \theta_2^2) + \gamma_2 \theta_1 \theta_2. \label{eq: lenstronomy_shear}
\end{equation}

Since the redshifts of the lens and source are known, the mass enclosed within the Einstein radius can be computed from the critical surface density as
\begin{equation}
M(<\theta_\mathrm{E}) = \pi \theta_{\mathrm{E}}^2 D_{\mathrm{d}}^2 \Sigma_{\mathrm{crit}} \; ,
\label{eq:mass}
\end{equation}

\noindent where

\begin{equation}
\Sigma_{\rm crit} = \frac{c^2}{4 \pi G} \frac{D_{\rm s}}{D_{\rm d} D_{\rm ds}},
\end{equation}
$c$ is the speed of light, $G$ is the gravitational constant, and $D_{\rm d}$, $D_{\rm s}$, and $D_{\rm ds}$ are the angular diameter distances to the lens, to the source, and between the lens and source, respectively.

The depth of the available data is insufficient to capture the radial slope of the lens mass distribution. 
The radial slope is highly correlated to inferred $H_0$ values \citep[e.g.,][]{Suyu2012b}. However, constraining the slope would require detailed modelling of the extended source surface brightness, a process that is computationally more expensive and, particularly, highly sensitive to the PSF reconstruction. As demonstrated by \cite{Shajib2022}, inaccuracies in the PSF model can introduce significant bias into these structural parameters. To avoid these systematic errors, we restrict our analysis to using the positions of the five SN images and constraining the following three mass models based on the standard singular profiles described above. Despite fixing the radial slope to isothermal, these models enable us to explore the level of complexity required in the lens mass distribution and to obtain estimates of the lensing magnifications. Following Occam's Razor in our sequence of models below, we start with the simplest form of mass models and increase complexity as required by the data.

\begin{enumerate}
    \item \textbf{Model I (SIS + SIS + shear):} Both G1 and G2 were modelled as SIS profiles. An external shear component was included to account for line-of-sight structures, environment, and effects arising from lack of model complexity \citep{10.1093/mnras/stae1375}.

    \item \textbf{Model II (SIE + SIS + shear):} G1 was modelled as an SIE while G2 was treated as an SIS. As in Model I, external shear was included.

    \item \textbf{Model III (SIE + SIE + shear):} G1 and G2 were both modelled as SIEs. As in the previous models, external shear was included.
\end{enumerate}

To mitigate the degeneracy between the lens position and external shear and to reduce the number of free parameters, the mass centroids were assigned Gaussian priors anchored to the lens light coordinates of the MGE fit in the $K$ band (Sect. \ref{sec: lenslightsub}; Table \ref{tab: astrometry}), as it is a good proxy of galaxy stellar mass \citep{Sureshkumar21}. Following SLACS results \citep{Bolton2008}, we adopt a width of $0.044\arcsec$, as the redshift of our system lies within the SLACS sample range. All other parameters were assigned uniform priors to ensure the model remains data-driven. A complete summary of the priors is listed in Table~\ref{tab:lensparams}.

\begin{table*}
\caption{Overview of lens components and priors of \texttt{lenstronomy} and \texttt{GLEE}.}
\label{tab:lensparams}
\centering
\renewcommand{\arraystretch}{1.4}
\begin{tabular*}{\textwidth}{l @{\extracolsep{\fill}} c c c}
\hline\hline
 & & \multicolumn{2}{c}{Priors} \\
\cline{3-4}
Component & Symbol (Unit) & \texttt{lenstronomy} & \texttt{GLEE} \\
\hline
\multicolumn{4}{l}{\textbf{G1 (SIE)}} \\
Centroid $x$ & $\theta_{1}$ ($''$) & $\mathcal{G}\left(\theta_{1,\rm G1}^{{\rm light}},0.044\right)$ & $\mathcal{G}\left(\theta_{1,\rm G1}^{{\rm light}},0.044\right)$ \\
Centroid $y$ & $\theta_{2}$ ($''$) & $\mathcal{G}\left(\theta_{2,\rm G1}^{{\rm light}},0.044\right)$ & $\mathcal{G}\left(\theta_{2,\rm G1}^{{\rm light}},0.044\right)$ \\
Einstein Radius & $\theta_{\rm E}$ ($''$) & $\mathcal{U}(0.5,3.0)$ & $\mathcal{U}(0.7,3.5)$ \\
Axis Ratio & $q$ & ... & $\mathcal{U}(0.5,1)$ \\
Position Angle & $\phi$ ($^\circ$) & ... & $\mathcal{U}(0,180)$ \\
Ellipticity & $\epsilon_1$ & $\mathcal{U}(-0.5,0.5)$ & ... \\
Ellipticity & $\epsilon_2$ & $\mathcal{U}(-0.5,0.5)$ & ... \\
\hline
\multicolumn{4}{l}{\textbf{G2 (SIS)}} \\
Centroid $x$ & $\theta_{1}$ ($''$) & $\mathcal{G}\left(\theta_{1,\rm G2}^{{{\rm light}}},0.044\right)$ & $\mathcal{G}\left(\theta_{1,\rm G2}^{{\rm light}},0.044\right)$ \\
Centroid $y$ & $\theta_{2}$ ($''$) & $\mathcal{G}\left(\theta_{2,\rm G2}^{{\rm light}},0.044\right)$ & $\mathcal{G}\left(\theta_{2,\rm G2}^{{\rm light}},0.044\right)$ \\
Einstein Radius & $\theta_{\rm E}$ ($''$) & $\mathcal{U}(0.5,2.5)$ & $\mathcal{U}(0.3,2.1)$ \\
\hline
\multicolumn{4}{l}{\textbf{External Shear}} \\
Magnitude & $\gamma_{\rm ext}$ & ... & $\mathcal{U}(0,1)$ \\
Angle & $\phi_{\rm ext}$ ($^\circ$) & ... & $\mathcal{U}(0,180)$ \\
Shear component & $\gamma_1$ & $\mathcal{U}(-0.5,0.5)$ & ... \\
Shear component & $\gamma_2$ & $\mathcal{U}(-0.5,0.5)$ & ... \\
\hline
\end{tabular*}
\tablefoot{
$\mathcal{U}(a,b)$ denotes a uniform prior between $a$ and $b$. $\mathcal{G}(c,d)$ denotes a Gaussian prior centred at $c$ with width $d$. \\ The superscripts $^{\rm light}$ refer to the centroid of the light profile of the respective galaxy (G1 or G2) in the $K$ band. \\ The conventions for ellipticity and external shear differ between the codes: \texttt{lenstronomy} adopts $\epsilon_{1,2}$ and $\gamma_{1,2}$, while \texttt{GLEE} uses $q$, $\phi$, $\gamma_{\rm ext}$ and $\phi_{\rm ext}$.
}
\end{table*}

\subsection{Evaluating lens models}
We discriminated between the models using the Bayesian Information Criterion \citep[BIC;][]{schwarz1978}, defined as
\begin{equation}
    \mathrm{BIC} \equiv \hat{\chi}^2_{\rm tot} + k \ln(n)
\end{equation}
where $\hat{\chi}^2_{\rm tot}$ is the minimum value of ${\chi}^2_{\rm tot}$ (goodness of fit), $k$ is the number of free parameters and $n$ is the number of constraints. The term, ${\chi}^2_{\rm tot}$, corresponds to a generalised negative log likelihood and explicitly accounts for penalties arising from the informative Gaussian priors:
\begin{equation}
    \chi^2_{\rm tot} = \chi^2_{\rm im} + \chi^2_{\rm prior}.
    \label{eq:chi2_tot}
\end{equation}
The expression for the first term quantifies the goodness of fit to the observed multiple image positions,
\begin{equation}
    \chi^2_{\text{im}} = \sum_{i=1}^{N_{\text{im}}} \frac{\left| \vec{\theta}_i^{\text{obs}} - \vec{\theta}_i^{~\text{mod}} \left( \vec{\eta} \right) \right|^2}{\sigma_i^2}, \label{eq: chi2_img}
\end{equation}
where $N_{\text{im}}$ is the number of multiple SN images, $\vec{\theta}_i^{\text{obs}}$ is the observed position of the $i$-th image, $\vec{\theta}_i^{~\text{mod}}$ is the model predicted position that depends on the values of the lens mass parameters $\vec{\eta}$, and ${\sigma_i^2}$ is the positional uncertainty associated with image $i$.  The second term is
\begin{equation}
  \chi^2_{\rm prior} = \sum_j \frac{(\eta_j - \mu_j)^2}{\sigma_{\eta,j}^2},
\label{eq: chi2_prior}
\end{equation}
where $\eta_j$ represents the value of the $j$-th model parameter with a Gaussian prior, while $\mu_j$ and $\sigma_{\eta,j}$ denote the mean and standard deviation of its Gaussian prior, respectively.

To calculate these metrics, we used the astrometric positions of the multiple images A--E as our primary observational constraints. While image flux can provide additional information, we excluded it from our analysis because it is susceptible to perturbations from millilensing and microlensing \citep[e.g.,][]{DalalKochanek2002, Nierenberg2017, DoblerKeeton2006, Huber+2019}. Additionally, the different SN images were captured at different phases, resulting in variations in intrinsic brightness that complicate their use as reliable constraints.

The above method was applied to both models \texttt{lenstronomy} and \texttt{GLEE}. The different modelling strategies are outlined in Sect.~\ref{sec:Lenstronomy} and \ref{sec:Glee}.

\subsection{Lenstronomy mass model}
\label{sec:Lenstronomy}
In \texttt{Lenstronomy}, the singular isothermal ellipsoid (SIE) is described by the projected surface mass density
\begin{equation}
    \kappa(x, y) = \frac{1}{2} \frac{\theta_{\rm E,Lenstronomy}}{\sqrt{q x^2 + y^2/q}},
\end{equation}
where $\theta_{\rm E,Lenstronomy}$ is the circularised Einstein radius, $q \leq 1$ is the minor-to-major axis ratio of the lens, and $x$ and $y$ are coordinates aligned with the major and minor axes of the lens. For G2, we set the axis ratio $q$ to one, which reduces it to a singular isothermal sphere.

Compared to the adapted SIE implementation in \citep{{Kassiola1993}} and described in Equation ~\ref{eq:piemd_convergence}, the circularised Einstein radius in \texttt{Lenstronomy} is related to the lens strength parameter $\theta_{\rm E}$ as
\begin{equation}
    \theta_{\rm E,Lenstronomy} = \frac{2 \sqrt{q}}{1 + q} \theta_{\rm E}.
\end{equation}
Similarly, \texttt{lenstronomy} implements shear using Cartesian components $(\gamma_1, \gamma_2)$ as described in eq. \eqref{eq: lenstronomy_shear}.

\subsubsection{Model Parameter optimisation}
For each of the $J$- and $K$-band images, we sampled the posterior probability distribution of the lens model parameters using a two-step optimisation and sampling strategy. First, a particle swarm optimisation (PSO; \citealt{Kennedy1995}) was employed to identify the global maximum of the likelihood and obtain robust initial values for the parameter inference. During this initial optimisation stage, we included an additional source-position penalising likelihood term, $\chi^2_\mathrm{scatter}$, which accelerates convergence by favouring compact source reconstructions and allows for a faster identification of suitable starting points in parameter space. Subsequently, the posterior distribution was explored using Markov chain Monte Carlo (MCMC) sampling with the affine-invariant ensemble sampler \texttt{emcee} \citep{ForemanMackey2013}.

The optimisation and sampling were performed by minimising the positional $\chi_{\rm im}^2$ in the image plane, defined in equation (\ref{eq: chi2_img}), summed with $\chi^2_\mathrm{scatter}$. We impose Gaussian priors on the mass centroids of G1 and G2, based on their light centroids (see Table \ref{tab:lensparams}).
During the sampling, \texttt{lenstronomy} was configured to enforce consistency with the observed image multiplicity by penalizing models that predict more than five images through an additional term in the log-likelihood.

\subsubsection{Predicted SN image positions and fluxes}
\label{sec:lenstronomy_pred}
Model~I ($k=10$ free parameters) was discarded because it could not reproduce the correct number of images. Calculating a BIC for this model is not meaningful, since we introduced an explicit penalising likelihood term for models producing more than five images, and this penalty was applied to every chain. The additional images predicted by Model~I have magnifications of $>10$, which would be detectable as demonstrated in Fig.~\ref{fig:detectability}. Even considering only the five matching images, the RMS scatter is the worst among the tested models, with an RMS of $\sim 0.05\arcsec$. For Models~II and III, the total $\chi^2$ is composed of three contributions: $\chi^2_\mathrm{scatter}$, which quantifies how well the source reconstruction is focused in the source plane (i.e., how tightly the multiple images map back to a common source position); $\chi^2_\mathrm{im}$ in equation (\ref{eq: chi2_img}), which measures how well the model reproduces the observed image positions; and $\chi^2_\mathrm{prior}$, which reflects the contribution from the applied parameter priors. For the $K$ band, Model~III ($k=14$) achieved a very low RMS of $\sim 0.0015\arcsec$; however, its total $\chi^2$ is $11.35$, composed of $\chi^2_\mathrm{scatter}\approx2.46$, $\chi^2_\mathrm{im}\approx0.62$, and $\chi^2_\mathrm{prior}\approx8.28$. Owing to the larger effective complexity, this results in a BIC of $43.6$ (for $n=10$ constraints in the $K$ band). Model~II ($k=12$) reaches a similarly low RMS of $\sim 0.0025\arcsec$ but with a slightly higher total $\chi^2$ of $11.62$, consisting of $\chi^2_\mathrm{scatter}\approx1.03$, $\chi^2_\mathrm{im}\approx1.99$, and $\chi^2_\mathrm{prior}\approx8.61$. This yields a BIC of $39.3$, providing positive evidence in its favor relative to Model~III ($\Delta \mathrm{BIC} \approx 4.3$; \citealt{Kass01061995}). The same conclusion is reached when evaluating the $J$ band independently. The posterior distributions of both models are shown in Fig.~\ref{fig:glee_corner_lenstronomy}. We note that the reported $\chi^2$ values and RMS correspond to the best-fit model, defined as the one minimizing the total $\chi^2$. While models with lower RMS or $\chi^2_{\mathrm{im}}$ do exist, they yield higher contributions from other $\chi^2$ terms and therefore result in a worse overall fit.

\begin{figure}[htb!]
    \centering
    \includegraphics[width=0.9\linewidth]{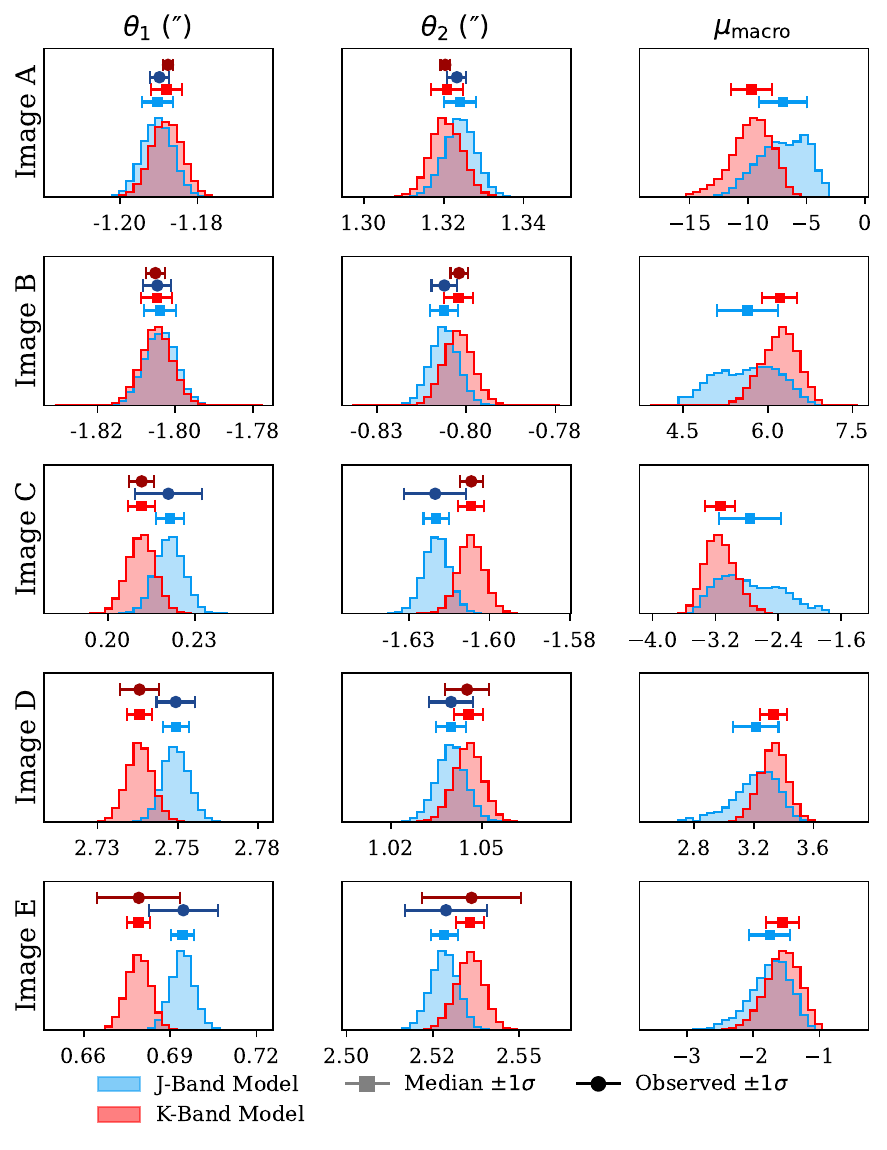}
    \caption{\texttt{Lenstronomy} posterior distributions of the lensed image positions ($\theta_1$, $\theta_2$) and magnifications for images A–E. Blue and red histograms correspond to the single band $J$ and $K$ models, respectively. Points above each histogram indicate the posterior mean, with horizontal error bars showing $\pm 1\sigma$ uncertainties.}
    \label{fig:lenstronomy_histograms_mb}
\end{figure}

The adopted Model~II reproduces the observed image positions with high fidelity (see Table~\ref{tab:lenstronomy_predictions}). The residuals of the predicted image positions are consistent with the astrometric precision estimated in Sect.~\ref{sec: astro+photo}. Figure~\ref{fig:lenstronomy_histograms_mb} shows a comparison between the posterior distributions of predicted image positions and magnifications.

\begin{table*}[tbp!]
\caption{\texttt{lenstronomy} model predicted astrometry and magnification of SN~Winny images A--E in the $J$ and $K$ bands. }
\label{tab:lenstronomy_predictions}
\centering
\renewcommand{\arraystretch}{1.25}
\begin{tabular}{l c c c c c c}
\hline\hline
\noalign{\smallskip}
 & \multicolumn{3}{c}{$J$ band} & \multicolumn{3}{c}{$K$ band} \\
Image & $\theta_1$ ($\arcsec$) & $\theta_2$ ($\arcsec$) & $\mu_{\rm macro}$ & $\theta_1$ ($\arcsec$) & $\theta_2$ ($\arcsec$) & $\mu_{\rm macro}$ \\
\noalign{\smallskip}
\hline
\noalign{\smallskip}
A & $-1.190 \pm 0.004$           & $\phantom{-}1.324 \pm 0.004$ & $\phantom{-}-6.89^{+2.03}_{-2.33}$    & $-1.188 \pm 0.004$           & $\phantom{-}1.321 \pm 0.004$ & $\phantom{-}-9.56^{+1.62}_{-1.83}$ \\
B & $-1.804 \pm 0.004$           & $-0.806 \pm 0.004$           & $\phantom{-}5.67^{+0.56}_{-0.64}$    & $-1.805 \pm 0.004$           & $-0.802 \pm 0.002$             & $\phantom{-}6.23^{+0.30}_{-0.35}$ \\
C & $\phantom{-}0.218 \pm 0.004$ & $-1.617 \pm 0.004$ & $\phantom{-}-2.81^{+0.49}_{-0.36}$    & $\phantom{-}0.210 \pm 0.004$ & $-1.606 \pm 0.004$ & $\phantom{-}-3.15^{+0.21}_{-0.18}$ \\
D & $\phantom{-}2.749 \pm 0.004$ & $\phantom{-}1.042 \pm 0.004$ & $\phantom{-}3.23^{+0.13}_{-0.17}$  & $\phantom{-}2.738 \pm 0.004$ & $\phantom{-}1.046 \pm 0.004$   & $\phantom{-}3.33 \pm 0.09$ \\
E & $\phantom{-}0.694 \pm 0.004$ & $\phantom{-}2.528 \pm 0.004$ & $\phantom{-}-1.72^{+0.27}_{-0.34}$    & $\phantom{-}0.679 \pm 0.004$ & $\phantom{-}2.536 \pm 0.004$   & $\phantom{-}-1.55^{+0.24}_{-0.27}$ \\
\noalign{\smallskip}
\hline
\end{tabular}
\tablefoot{Positions are measured relative to the mean centre of light of G1 in the $K$ band, with $\theta_1$ oriented West and $\theta_2$ North. The constraints on $J$ and $K$ bands are obtained through separate single-band modellings.}
\end{table*}

\subsection{\texttt{GLEE} Mass model}
\label{sec:Glee}
We modelled this system with \texttt{GLEE}. We adopted the same mass configurations described in Sect. \ref{sec: massparam}. For each scenario, the free parameters of the mass profiles correspond to the Einstein radii and structural parameters of deflectors G1 and G2 (see Table \ref{tab:lensparams}), along with the external shear magnitude and direction.

\subsubsection{Model parameter optimisation}
We sampled the posterior probability distribution of the lens parameters using a combination of simulated annealing and MCMC. We optimised the fit by minimising the positional $\chi_{\rm im}^2$ in the image plane, defined in equation \eqref{eq: chi2_img}, while imposing Gaussian priors on the lens mass centroids based on their light centroids (Table \ref{tab:lensparams}).

We utilised the lensed SN image positions and their uncertainties in Table \ref{tab: astrometry}.
Furthermore, we enforced a null-detection constraint to ensure consistency with the observed image multiplicity. We explicitly discarded any model realisation predicting additional images that were not observed, provided their predicted macro-magnification exceeded $|\mu_{\rm macro}| > 0.1$.

We calibrated this exclusion threshold using synthetic data, in which we injected a PSF over a Gaussian noise distribution comparable to that of the science data. These tests, shown in Fig.~\ref{fig:detectability}, demonstrate that a magnification of $\approx 0.4$ yields a peak signal-to-noise ratio (SNR) of only $\sim 1\sigma$. At this level, the signal is visually indistinguishable from the background noise. Nevertheless, we retained the more conservative exclusion threshold of $|\mu_{\rm macro}| > 0.1$. This factor of $\sim 4$ difference provides a safety margin accounting for the observed variations in the amplitude to model predicted macro-magnification ratio ($A/ |\mu_{\rm macro}|$, shown in Sect. \ref{sec:flux_ratio}) across the observed images (A--E) due to potential microlensing and millilensing magnification. This threshold of $|\mu_{\rm macro}| > 0.1$ ensures that we do not accept models which produce detectable additional images, although in practice, the results remain the same irrespective of threshold values ranging from 0.1 to 1.5 given that the predicted macro-magnifications of additional images are higher than the range of threshold values (shown in \ref{sec: glee_pred}).

The resulting posterior probability distributions for the free parameters of our best-fitting Model~II are shown in Appendix Fig.~\ref{fig:glee_corner_mb}. These demonstrate that the parameters are well-constrained and free of significant degeneracies.

\subsubsection{Predicted SN image positions and fluxes}
\label{sec: glee_pred}

In the \texttt{GLEE} multiband models, the parameter count $k$ includes two additional degrees of freedom relative to single-band \texttt{lenstronomy} fits, corresponding to the source position in the second band. Model~I ($k=12$) was rejected as it fully failed to reproduce the image multiplicity and predicted a mass centroid offset of up to $0.3\arcsec$ from the stellar light. With an RMS between observed and predicted image positions of ${\approx0.0099\,\arcsec}$, a ${\,\chi^2_{\rm im} \approx 16.7}$, and a substantial prior penalty of ${\chi^2_{\rm prior} \approx 87.4}$, it resulted in a BIC of $140.0$. 

We adopted Model~II ($k=14$) as the preferred solution. It yielded an ${{\rm RMS}\approx0.0071\arcsec}$, a ${\,\chi^2_{\rm im} \approx 11.2}$, and $\chi^2_{\rm prior} \approx 8.0$, resulting in a BIC of $61.1$. While Model~III ($k=16$) achieved marginally lower astrometric residuals (${\rm RMS}\approx 0.0069\arcsec$, $\chi^2_{\rm im} \approx 10.7$) with a comparable prior penalty ($\chi^2_{\rm prior} \approx 7.8$), this came at the cost of increased complexity. The resulting BIC of $66.4$ provides positive evidence ($\Delta \mathrm{BIC} \approx 5.3$) in favour of the simpler Model~II \citep{Kass01061995}.

\begin{figure}[htb!]
    \centering
    \includegraphics[width=0.9\linewidth]{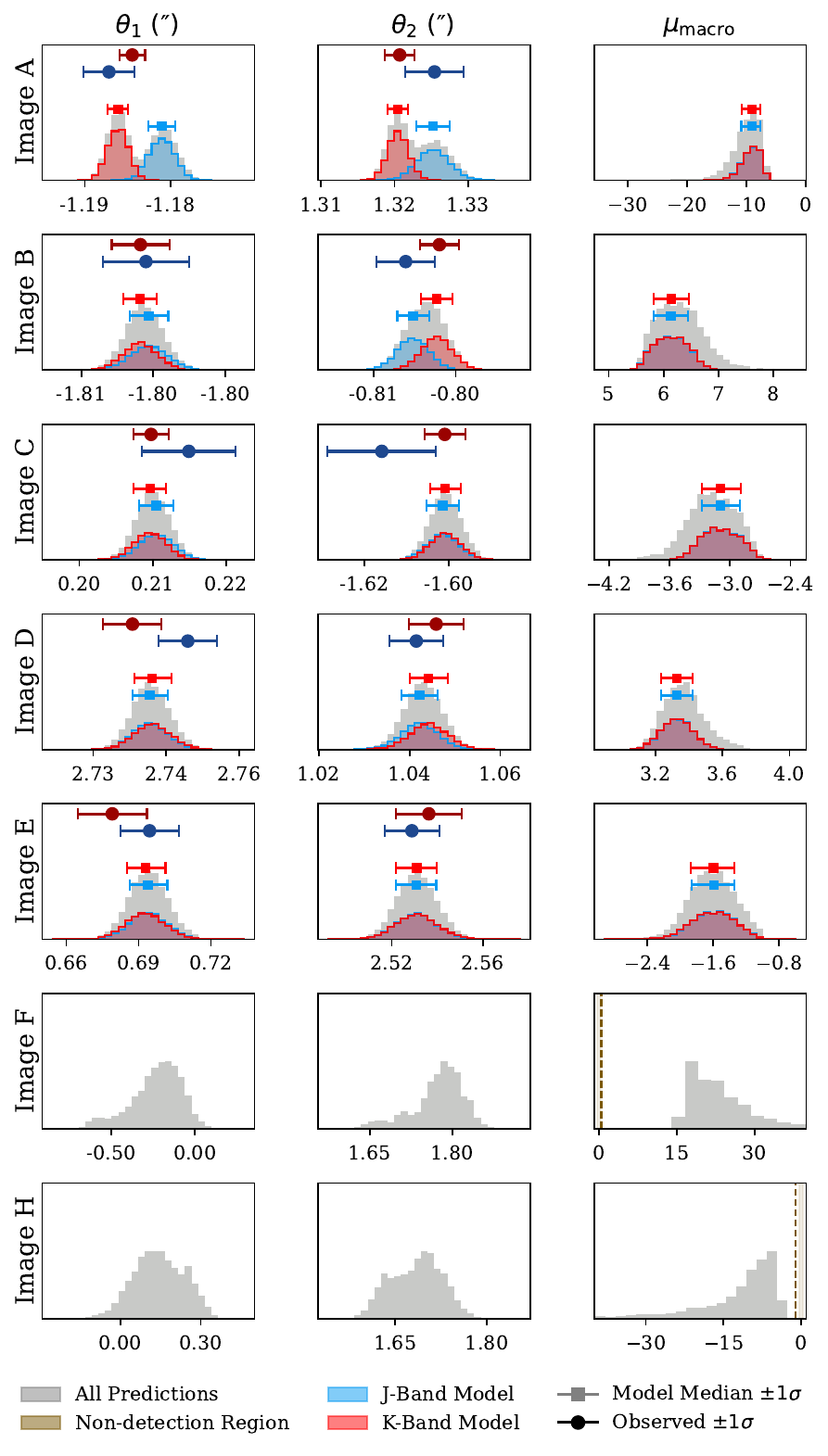}
    \caption{Posterior distributions of the predicted image positions ($\theta_{\rm 1}$, $\theta_{\rm 2}$) and macro-model magnifications ($\mu_{\rm macro}$) for the multiband \texttt{GLEE} model (Model~II). The grey histograms show the full distribution of all model predictions, using the five SN images as constraints without penalising models that predict additional images. The coloured histograms represent the subsets of models that validly reproduce the observed image multiplicity (=5) in the $J$ band (blue) and $K$ band (red). The lighter error bars with a square marker indicate the median and $1\sigma$ intervals for these subsets, while the darker error bars with a circle marker show the observed astrometry with $1\sigma$ uncertainties. The bottom rows display the distributions for the additional counter-images F and H, which are predicted to be magnified ($|\mu_{\rm macro}| > 0.1$) but are not detected in the observations. The region $|\mu_{\rm macro}| \leq 0.1$, where an image may be present but obscured by noise, is highlighted in brown.}
    \label{fig:glee_histograms_mb}
\end{figure}

The preferred Model~II reproduces the observed image positions with high accuracy (see Table~\ref{tab:glee_predictions}). We find that the image position residuals are consistent with the astrometric precision derived in Sect.~\ref{sec: astro+photo} across both bands. Figure~\ref{fig:glee_histograms_mb} compares the predicted posterior distributions of the image positions and magnifications against the observed values, showing excellent agreement across all five observed images (A--E).

In addition to the multiband analysis, we constructed a single-band variant (Model~II) using only the $K$-band image positions to facilitate a direct comparison with the \texttt{lenstronomy} results. With $k=12$ free parameters constrained by only $n=10$ observables, this geometric configuration is formally under-determined; however, the informative Gaussian priors help to alleviate this, but not fully. The model fully recovers the astrometry, yielding an ${\rm RMS}=0.0012\,\arcsec$, ${\chi^2_{\rm im}=0.07}$,~${\chi^2_{\rm prior}=6.43}$, and BIC=34.13. We caution that model comparison using BIC values are only meaningful when the data set is the same for the models.

While the resulting parameter estimates remain highly consistent with the primary multiband model, the lack of sufficient constraints leads to significant degeneracies and the irregular posterior distributions shown in Appendix~\ref{sec: glee_sb}. This behaviour shows that while a single-band position-only model can accurately recover the lens parameters, the additional information from multiband data is essential to break these degeneracies and achieve a well-constrained, unique solution. It is important to note that while the \texttt{lenstronomy} analysis incorporates additional likelihood terms (a source position penalty) to stabilise the fit, our Model~II relies strictly on image positions.

\begin{table*}[tbp!]
\caption{Same as Table \ref{tab:lenstronomy_predictions} but for the \texttt{GLEE} multiband mass model.}
\label{tab:glee_predictions}
\centering
\renewcommand{\arraystretch}{1.25}
\begin{tabular}{l c c c c c c}
\hline\hline
\noalign{\smallskip}
 & \multicolumn{3}{c}{$J$ band} & \multicolumn{3}{c}{$K$ band} \\
Image & $\theta_1$ ($\arcsec$) & $\theta_2$ ($\arcsec$) & $ \mu_{\rm macro} $ & $\theta_1$ ($\arcsec$) & $\theta_2$ ($\arcsec$) & $\mu_{\rm macro}$ \\
\noalign{\smallskip}
\hline
\noalign{\smallskip}
A & $-1.185 \pm 0.001$ & $\phantom{-}1.323 \pm 0.001$ & $-9.1^{+1.4}_{-1.8}$        & $-1.189 \pm 0.001$ & $\phantom{-}1.320 \pm 0.001$ & $-9.0^{+1.4}_{-1.8}$ \\
B & $-1.804 \pm 0.002$ & $-0.805 \pm 0.002$           & $\phantom{-}6.13 \pm 0.32$  & $-1.805 \pm 0.001$ & $-0.802 \pm 0.002$           & $\phantom{-}6.1 \pm 0.3$ \\
C & $\phantom{-}0.211 \pm 0.004$ & $-1.606 \pm 0.003$ & $-3.09^{+0.20}_{-0.19}$     & $\phantom{-}0.209 \pm 0.003$ & $-1.606 \pm 0.003$           & $-3.1\pm0.2$ \\
D & $\phantom{-}2.742 \pm 0.004$ & $\phantom{-}1.042 \pm 0.004$ & $\phantom{-}3.33 \pm 0.09$  & $\phantom{-}2.742 \pm 0.004$ & $\phantom{-}1.044 \pm 0.004$ & $\phantom{-}3.3 \pm 0.1$ \\
E & $\phantom{-}0.694 \pm 0.008$ & $\phantom{-}2.531 \pm 0.009$ & $-1.59^{+0.25}_{-0.27}$     & $\phantom{-}0.693 \pm 0.008$ & $\phantom{-}2.531 \pm 0.009$ & $-1.6\pm 0.3$ \\
\noalign{\smallskip}
\hline
\end{tabular}
\end{table*}

\section{Discussion}\label{sec:discussion}
We compare the lens models obtained using two independent lens modelling codes (namely \texttt{lenstronomy} and \texttt{GLEE}). Despite differences in numerical implementation and optimisation strategies, the two approaches yield broadly consistent results. As discussed in Sect.~\ref{sec:lenstronomy_pred} and Sect.~\ref{sec: glee_pred}, Model~II was identified as the best-performing model according to the BIC criterion. A single-band $K$-band model is used as a reference to compare the two models.

\subsection{Comparison of Lenstronomy and GLEE model results}

Image-plane constraints: Both models reproduce the observed image positions with high accuracy. To quantify their mutual consistency, the RMS separation between the predicted image positions of the two models is computed, yielding a value of $0.0027\, \arcsec$. This is approximately a factor of 5.5 smaller than the pixel scale, demonstrating that the two models are fully consistent with one another.

For each parameter $i$, the level of agreement was quantified as
\begin{equation}
N_{\sigma,i} =
\frac{|\eta_{{\rm Lenstronomy},i} - \eta_{{\rm GLEE},i}|}
{\sqrt{\sigma_{\eta,{\rm Lenstronomy},i}^2 + \sigma_{\eta,{\rm GLEE},i}^2}},
\end{equation}
where $\eta_{{\rm Lenstronomy},i}$ and $\eta_{{\rm GLEE},i}$ denote the lens parameter from the two models, and $\sigma_{\eta,{\rm Lenstronomy},i}$ and $\sigma_{\eta,{\rm GLEE},i}$ are the corresponding (symmetrised) uncertainties. All parameters yield $N_\sigma < 1$, meaning the differences between the two models are smaller than the combined $1\sigma$ uncertainties. This indicates that the inferred lens parameters and magnifications are fully consistent, with no significant tension.

\begin{table*}
\caption{Comparison of lens model parameters obtained with \texttt{lenstronomy} and \texttt{GLEE} for the single $K$-band modelling. Uncertainties are quoted as $1\sigma$ confidence intervals.}
\label{tab:lens_results_comparison}
\centering
\renewcommand{\arraystretch}{1.3}
\begin{tabular*}{\textwidth}{l @{\extracolsep{\fill}} c c c}
\hline\hline
Component & Symbol (Unit) & \texttt{lenstronomy} & \texttt{GLEE} \\
\hline
\multicolumn{4}{l}{\textbf{G1 (SIE)}} \\
Centroid $x$ & $\theta_{1}$ ($''$) & $-0.058_{-0.009}^{+0.009}$ & $-0.057_{-0.007}^{+0.009}$ \\
Centroid $y$ & $\theta_{2}$ ($''$) & $-0.094_{-0.004}^{+0.004}$ & $-0.094_{-0.004}^{+0.005}$ \\
Einstein Radius & $\theta_{\rm E}$ ($''$) & $1.609_{-0.013}^{+0.009}$ & $1.607_{-0.008}^{+0.010}$ \\
Axis Ratio & $q$ & $0.74_{-0.02}^{+0.02}$ & $0.74_{-0.02}^{+0.01}$ \\
Position Angle & $\phi$ ($^\circ$) & $52_{-5}^{+5}$ & $53_{-4}^{+5}$ \\
\hline
\multicolumn{4}{l}{\textbf{G2 (SIS)}} \\
Centroid $x$ & $\theta_{1}$ ($''$) & $1.01_{-0.03}^{+0.04}$ & $1.01_{-0.04}^{+0.03}$ \\
Centroid $y$ & $\theta_{2}$ ($''$) & $2.17_{-0.03}^{+0.03}$ & $2.16_{-0.02}^{+0.03}$ \\
Einstein Radius & $\theta_{\rm E}$ ($''$) & $0.745_{-0.008}^{+0.007}$ & $0.743_{-0.010}^{+0.010}$ \\
\hline
\multicolumn{4}{l}{\textbf{External Shear}} \\
Magnitude & $\gamma_{\rm ext}$ & $0.114_{-0.006}^{+0.008}$ & $0.116_{-0.005}^{+0.005}$ \\
Angle & $\phi_{\rm ext}$ ($^\circ$) & $25_{-2}^{+2}$ & $24_{-2}^{+2}$ \\
\hline
\end{tabular*}
\end{table*}

The magnifications predicted by the two lens models were compared using a parameter-difference significance test, analogous to that applied to the lens model parameters. All images are consistent within $\lesssim 1\sigma$, with $N_\sigma$ values of 0.18, 0.24, 0, 0.17, 0 for images A through E, respectively. This demonstrates that the two models predict magnifications that are statistically compatible. Such consistency is expected, as magnifications are sensitive to small-scale variations in the lens potential, but the global agreement confirms that both models reproduce the lensing configuration accurately.

    \begin{table}
    \centering
    \caption{Comparison of predicted magnifications between the two models single K-band models.
    }
        \begin{tabular}{cccc}
        \hline
        Image & $\mu_{\rm lenstronomy}$ & $\mu_{\rm GLEE}$ & $N_\sigma$ \\
        \hline
        A & $-9.6 \pm 1.7$  & $-10.0 \pm 1.5$ & 0.18 \\
        B & $6.2 \pm 0.3$   & $6.3 \pm 0.3$   & 0.24 \\
        C & $-3.2 \pm 0.2$  & $-3.2 \pm 0.2$  & 0 \\
        D & $3.33 \pm 0.09$ & $3.35 \pm 0.08$ & 0.17 \\
        E & $-1.6 \pm 0.3$  & $-1.6 \pm 0.3$  & 0 \\
        \hline
        \end{tabular}
        \tablefoot{The last column indicates the significance of the difference in units of the
    combined uncertainty.}
        \end{table}
\noindent

\subsection{Mass does not strictly follow light}
\begin{figure}[htb!]
    \centering
    \includegraphics[width=1\linewidth]{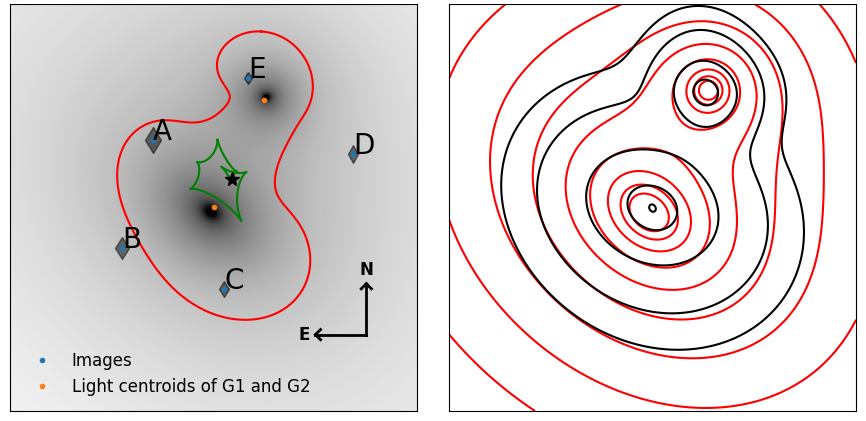}
    \caption{Left: Convergence map for SN Winny obtained with Model~II for \texttt{lenstronomy} in the $K$ band. The critical curves (red) and caustics (green) are shown together with the reconstructed source position (black star) and the observed multiple-image positions (dots). The predicted image positions from the lens model are indicated by diamonds, with their sizes reflecting the magnifications. Right: Contour comparison between the projected surface mass density ($\kappa$; red contours) and the luminous mass model derived from the MGE light profile (black contours), shown on the same angular scale.}
    \label{fig:contours}
\end{figure}

All models were assigned a Gaussian prior on the mass centroid, based on the
averaged position of the two MGE models used to describe the light distribution,
with a width of $0.044\arcsec$. Despite this informative prior, the inferred
position of the mass consistently deviates from the light centroid, indicating
that the mass distribution does not perfectly trace the observed stellar light.
The distances between the inferred mass centroids and the light centroids
were calculated for each lens component, yielding offsets of
$d_\mathrm{G1} = 0.112\arcsec$ and $d_\mathrm{G2} = 0.065\arcsec$
(corresponding to 7.5 and 4.3 pixels, or 0.58 kpc and 0.34kpc respectively). These offsets are significantly larger than the width of the Gaussian prior and the uncertainty of the image position from the lens light modelling, indicating that the mass does not follow light. Even after including additional model complexity, such as external shear, the offsets persist, demonstrating that the mass-light misalignment is robust. Given the presence of two galaxies G1 and G2 that may share a common dark matter halo or are interacting, it is not surprising that mass does not strictly follow light in this binary lens system. This offset can be seen in Fig.~\ref{fig:contours}.
Using Eq.~\ref{eq:mass}, we measure the enclosed mass within the Einstein radius as $M_{\rm G1}(<\theta_{\rm E}) = 4.61^{+0.06}_{-0.04} \times 10^{11}\,M_\odot$ for G1 and $M_{\rm G2}(<\theta_{\rm E}) = 1.01_{-0.02}^{+0.02} \times 10^{11}\,M_\odot$ for G2. The DESI collaboration \citep{desicollaboration2025datarelease1dark} reports a total stellar mass estimate for G1 of 
$\log(M_\ast/M_\odot)=11.55$ derived from \citet{fast_spec2023}, using a Chabrier IMF, $H_0=100\,\mathrm{km\,s^{-1}\,Mpc^{-1}}$ 
and an effective radius of $2.65\,\arcsec$. Scaling these values to $H_0=70\,\mathrm{km\,s^{-1}\,Mpc^{-1}}$, a Salpeter IMF, 
and extrapolating to the Einstein radius yields a stellar mass fraction of $\sim 0.48$, consistent with the stellar mass 
fractions inferred within the Einstein radii of massive early-type galaxies in the SLACS sample \citep{Slacs_2009}.

\subsection{Flux-ratio anomaly}
\label{sec:flux_ratio}
The amplitudes measured by fitting the PSF to the lens-light-subtracted images were compared to the magnification predictions from both \texttt{GLEE} and \texttt{lenstronomy}. As shown in Fig.~\ref{fig: colour}, the colours of the supernova images are broadly consistent, though small differences remain.

\begin{figure*}
\centering
\includegraphics[width=0.8\linewidth]{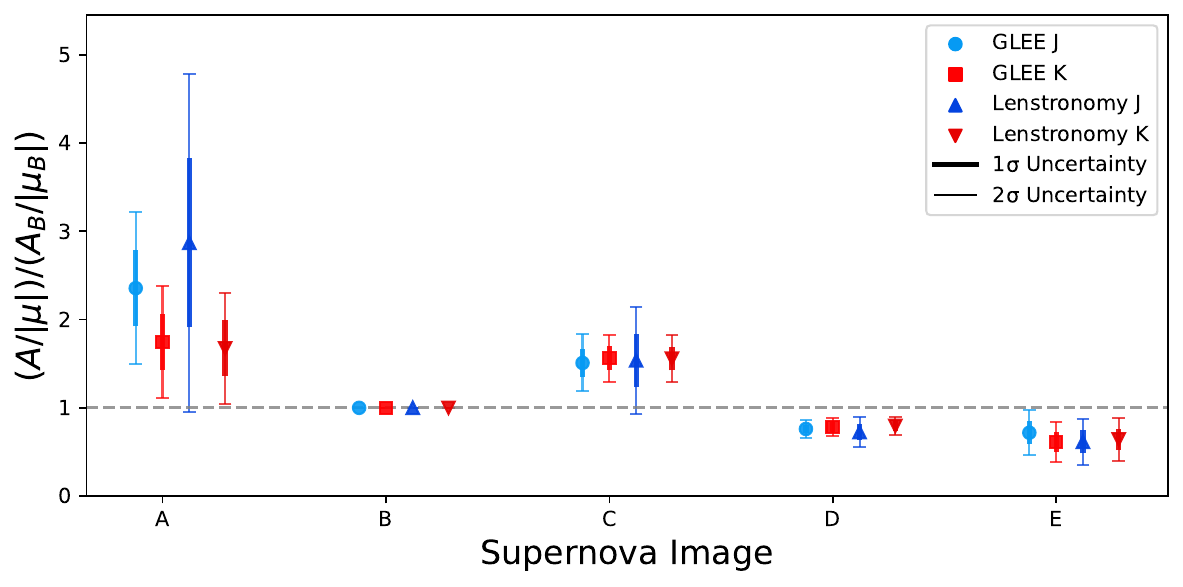}
\caption{Normalised amplitude-to-macro-model magnification ratios ($A / \mu_\mathrm{macro}$) for the five lensed images of SN~Winny. Blue and red markers correspond to the $J$ and $K$ bands, respectively. Circles and squares show comparisons using \texttt{GLEE} predictions, while triangles and inverted triangles show comparisons using \texttt{lenstronomy} predictions. The thick and thin error bars indicate $1\sigma$ and $2\sigma$ uncertainties, respectively. Image B is normalised to 1 in both bands. Observed amplitudes from PSF fitting were used for both models.}
\label{fig:AtoMu}
\end{figure*}

To enable a direct comparison, we normalised the amplitude-to-magnification ratios by the ratio of image B for each of the $K$ and $J$ bands such that image B has a value of 1. The resulting normalised ratios are displayed in Fig.~\ref{fig:AtoMu}. By construction, image B matches perfectly. Across all five images, the normalised amplitude-to-magnification ratios from both models do not agree within $2\sigma$ (except for the $J$ band predictions of \texttt{lenstronomy} for SN images A and C given their larger uncertainties).  The relative flux ratios predicted by the macro models of \texttt{GLEE} and \texttt{lenstronomy} are thus mostly inconsistent with the observed amplitudes.

Several factors could contribute to this discrepancy: (1) microlensing by a compact object, such as a star, black hole, globular cluster (can in extreme cases add up to $\Delta m = -2.5 \log_{10} (\mu / \langle \mu_{\rm macro} \rangle) \sim 1.5$\,mag as shown by \citealt{weisenbach_2021}); (2) millilensing by dark matter substructure \citep[e.g.,][]{DalalKochanek2002, Nierenberg2017}; (3) the supernova images are observed at different times due to lensing time delays, so intrinsic brightness variations of the source can lead to mismatches between observed amplitude and predicted magnification; (4) the simplicity of the adopted macro model, which consists of two singular isothermal profiles constrained only by image positions, may not capture the full mass distribution; (5) contamination from the host galaxy light. Incorporating the host galaxy arcs from future data sets or adopting more complex mass profiles could improve the model and potentially change the predicted magnifications.

\subsection{Fifth image}
To further test the robustness of the lensing configuration, we constructed an additional lens model using only the four securely identified supernova images as positional constraints, explicitly excluding the candidate fifth image from the modelling procedure. This approach allows us to assess whether the presence and location of the fifth image are naturally predicted by the mass configuration inferred from the remaining images.

Using the same mass parameterisation as in the preferred model (SIE for G1, SIS for G2, and external shear), the optimisation converges to a solution that reproduces the four input image positions. Despite not being included as a constraint, the model predicts the formation of a fifth image in close proximity to G2.

The predicted position of this image is offset from the observed candidate location by a distance of $0.49\, \arcsec$. Given that the number of free model parameters significantly exceeds the number of positional constraints provided by only four images, the system is inherently underconstrained. In such a regime, larger positional residuals are expected, particularly for images that are not explicitly included in the fit. 

Nevertheless, it is noteworthy that a fifth image is generically produced by the model and appears near the secondary lens galaxy G2, consistent with the location inferred from the lens-light-subtracted imaging. This provides additional qualitative support for the interpretation of the system as a five-image strong lens configuration. 

Lastly, we highlight that immediately following the detection of SN~Winny, initial lens models we generated using the archival CFHT data mentioned in \citet{Taubenberger2025} had already predicted a fifth image at the location currently identified in the LBT data and the difference imaging from newer CFHT observations \citep{Aryan+2025}.

\section{Summary and Outlook}

We present the first mass model of the lensed SN Winny system, confirming the presence of a suspected fifth image near the secondary lens galaxy G2. Both lens galaxies, G1 and G2, lie at a common redshift, enabling a consistent two-deflector modelling approach. Our analysis uses deep $J$- and $K$-band imaging from the LBT, in which the lens galaxy light was modeled down to the noise level and the positions of the supernova images were measured with high precision. These positions served as constraints for independent lens modelling with \texttt{lenstronomy} and \texttt{GLEE}.

In addition to the arguments presented by \citet{Aryan+2025}, we provide two further pieces of evidence that the system indeed consists of five images. First, the colour of the candidate fifth image is consistent with the other supernova images across the $J$ and $K$ bands, supporting a common origin. Second, even when image E is excluded from the modelling constraints, the best-fitting lens model still predicts the formation of a fifth image near G2. .

Of the three classes of isothermal lens models we explored, the preferred mass model consists of a SIE for G1, a SIS for G2, and an external shear component, selected based on the BIC. Despite differences in implementation, both modelling software and frameworks produce consistent results across both filters. Our final models do not produce additional SN images that are not detected.

From the lens models, we measure the enclosed mass within the Einstein radius as $M_{\rm G1}(<\theta_{\rm E}) = 4.61^{+0.06}_{-0.04} \times 10^{11}\,M_\odot$ for G1 and $M_{\rm G2}(<\theta_{\rm E}) = 1.01_{-0.02}^{+0.02} \times 10^{11}\,M_\odot$ for G2. This is consistent with G1 being an elliptical galaxy and G2 a lower mass companion. The observed relative fluxes of most of the SN images are inconsistent (at the $\gtrsim2\sigma$ level but within $3\sigma$) with that of the model-predicted fluxes; this may indicate microlensing by a compact object, a scenario that can be tested once well-sampled light curves become available. Observations from facilities such as the COLIBRI \citep{colibri_2022}, LOT, Maidanak \citep{Maidanak_2018} and Wendelstein telescope \citep{hopp_2014,Lang-Bardl_2016} are currently monitoring this system to provide the necessary time-domain information \citep{Taubenberger2025}.

Time-delay measurements combined with cosmography-optimised strong lens models enable an inference of the Hubble constant. Our mass models, which fit well the observed SN image positions with an RMS in the observed and predicted image positions of $\sim 0.0012\, \arcsec - 0.0025\, \arcsec$, show that this system is promising for cosmographic constraints. Since the inferred $H_0$ value depends on the radial lens mass profile, which our current data do not allow us to constrain, future data showing clearly the Einstein ring are required to constrain the radial lens mass profile. For this reason, we refrain from making time-delay predictions and cosmographic forecasts in this work. Nonetheless, our mass model forms the basis for future cosmographic-grade mass models.

SN Winny is the first galaxy-scale lens system that is promising to yield an $H_0$ measurement with $\lesssim$10\% uncertainty. Until now, such cosmographically useful lensed SNe are rare on galaxy-scale.  Nonetheless, with Rubin Observatory Legacy Survey of Space and Time providing time-domain triggers for supernovae and \textit{Euclid} confirming these systems as strong lenses while delivering initial lens models, we anticipate dozens of lensed SNe per year \citep[e.g.,][]{OguriMarshall2010, Wojtak+2019, SainzdeMurieta+2024, Arendse+2024}. The combination of both surveys will allow the Hubble constant to be determined with high precision from an ensemble of such systems.

\begin{acknowledgements}
We thank the staff of the LBT observatory for their support during the execution of the observations. The LBT is an international collaboration
among institutions in the United States, Italy and Germany. LBT
Corporation partners are: The University of Arizona on behalf of the
Arizona university system; Istituto Nazionale di Astrofisica, Italy;
LBT Beteiligungsgesellschaft, Germany, representing the Max-Planck
Society, the Astrophysical Institute Potsdam, and Heidelberg
University; The Ohio State University, and The Research Corporation,
on behalf of The University of Notre Dame, University of Minnesota
and University of Virginia.

AGS, LD, SHS and EM thank the Max Planck Society for support through the Max Planck Fellowship for SHS. 
LD acknowledges support from the China Scholarship Council.
This work is supported in part by the Deutsche Forschungsgemeinschaft (DFG, German Research Foundation) under Germany's Excellence Strategy -- EXC-2094 -- 390783311.

RC acknowledges support from the French government under the France 2030 investment plan, as part of the Initiative d'Excellence d'Aix-Marseille Universit\'e -- A*MIDEX (AMX-23-CEI-088).

T.-W.C. acknowledges financial support from the Yushan
Fellow Program of the Ministry of Education, Taiwan
(MOE-111-YSFMS-0008-001-P1), and from the National Science and Technology
Council, Taiwan (NSTC 114-2112-M-008-021-MY3).

AG acknowledges funding and support by the Swiss National Science Foundation (SNSF).

This work made use of \texttt{Astropy}: a community-developed core Python package and an ecosystem of tools and resources for astronomy \citep{astropy13,astropy18,astropy22}, \texttt{NumPy} \citep{numpy}, \texttt{Matplotlib} \citep{matplotlib}, \texttt{pandas} \citep{pandas}, \texttt{corner} \citep{corner}, and \texttt{emcee} \citep{ForemanMackey2013}.
\end{acknowledgements}

\bibliographystyle{./bibtex/aa} 
\bibliography{bib} 

\begin{appendix}

\onecolumn

\section{Detectability of images}

\begin{figure}[hbt]
    \centering
    \includegraphics[width=\linewidth]{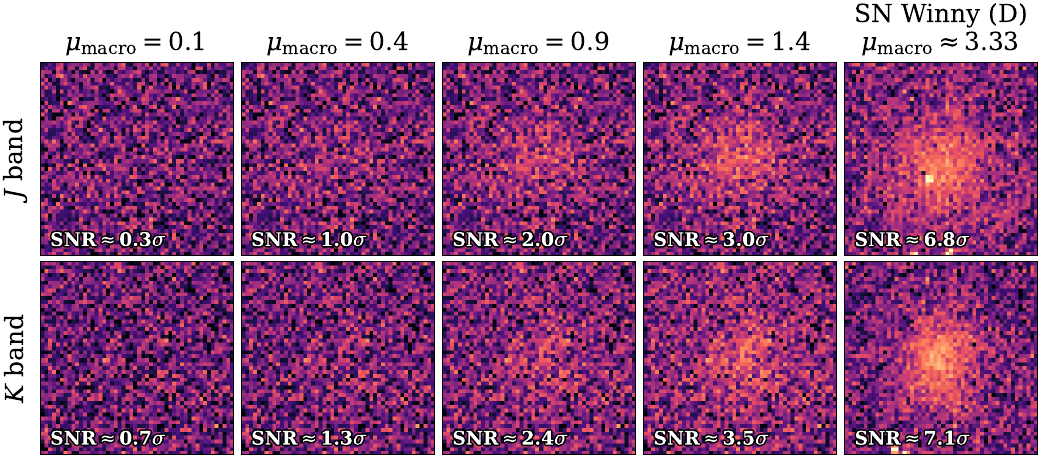}
    \caption{Visual detectability of point sources as a function of predicted magnification assuming no microlensing nor millilensing. The top and bottom rows display $J$-band and $K$-band cutouts ($50 \times 50$ pixels), respectively. The first four columns show synthetic point source injections into Gaussian noise fields with statistics ($\sigma_{\rm bkg}$) matched to the science data, simulating magnifications of $\mu_{\rm macro}=0.1, 0.4, 0.9,$ and $1.4$. The rightmost column shows the observed science data for SN~Winny (Image D). The annotated values indicate the peak signal-to-noise ratio ($\sigma$). Note that even at $\mu_{\rm macro}=0.4$, the signal remains consistent with the background noise ($\sim 1\sigma$) across all pixels, demonstrating that any model prediction with $\mu_{\rm macro} \lesssim 0.4$ would be visually undetectable (unless the lensing magnification is altered by millilensing or microlensing).} 
    \label{fig:detectability}
\end{figure}
\FloatBarrier

\newpage

\section{\texttt{lenstronomy} Single-band Model - Parameter Distribution}
\begin{figure*}[h!]
    \centering
    \includegraphics[width=\linewidth]{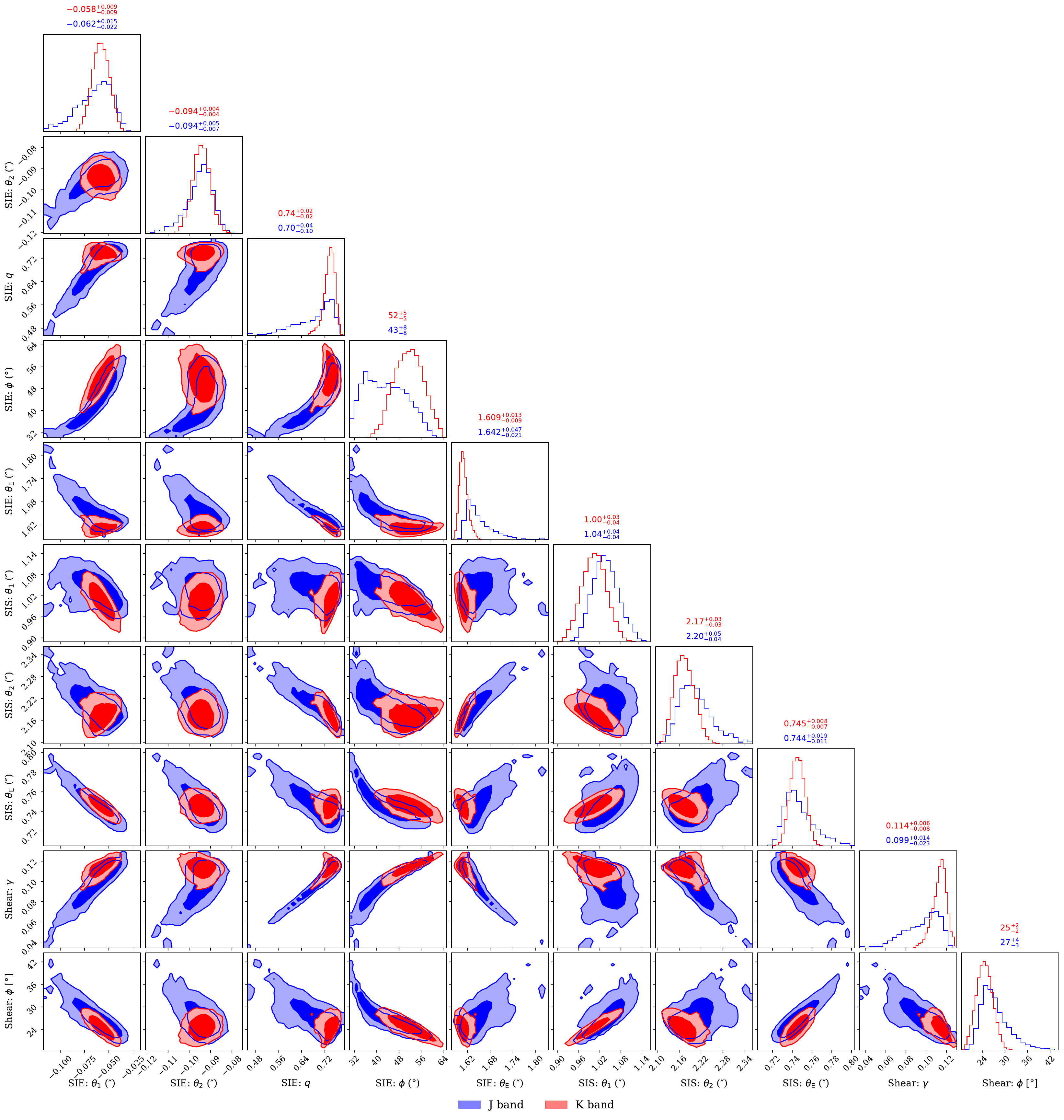}
    \caption{Posterior distributions of the lens model parameters for the best-fitting \texttt{lenstronomy} model (Model~II).  The blue and red histograms represent the $J$ and $K$-band models respectively. The numerical values above the diagonal columns indicate the median and 1$\sigma$, while the off-diagonal panels display the 2D covariances between the model parameters.}\label{fig:glee_corner_lenstronomy}
\end{figure*}
\FloatBarrier

\newpage

\section{\texttt{GLEE} Multiband Model - Parameter Distribution}

\begin{figure*}[h!]
    \centering
    \includegraphics[width=\linewidth]{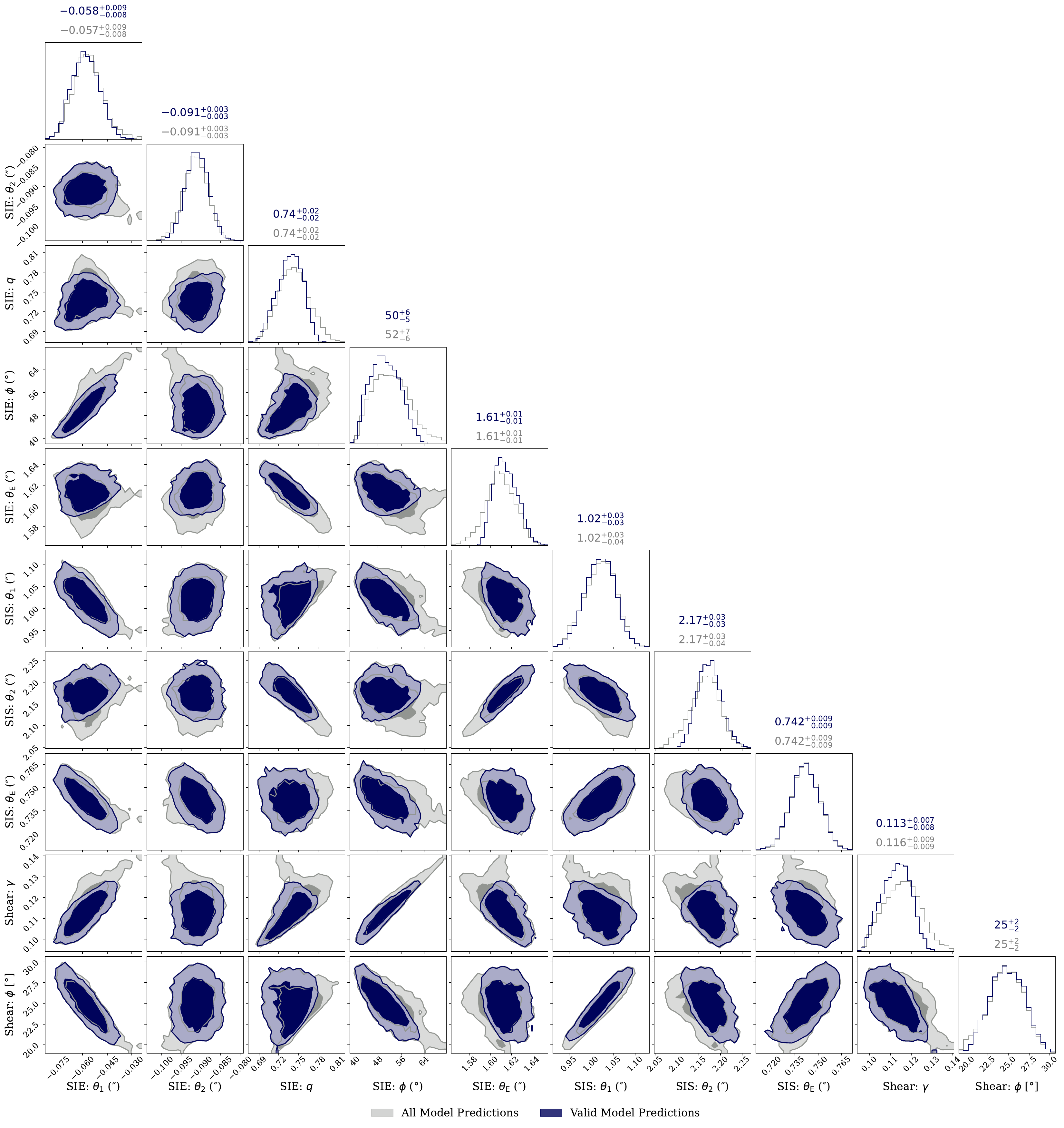}
    \caption{Posterior distributions of the lens model parameters for the best-fitting \texttt{GLEE} model (Model~II). The grey histograms show the full distribution of all model predictions. The coloured histograms (blue) represent the subset of models that validly reproduce the observed image multiplicity. The numerical values above the diagonal columns indicate the median and 1$\sigma$, while the off-diagonal panels display the 2D covariances between the model parameters.}\label{fig:glee_corner_mb}
\end{figure*}
\FloatBarrier

\newpage

\section{\texttt{GLEE} Single $K$ Band Model} \label{sec: glee_sb}

\begin{table*}[h!]
\caption{\texttt{GLEE} single band model predicted astrometry and macro-magnification $\mu_{\rm macro}$ of SN~Winny images A--E. }
\label{tab:glee_predictions_single}
\centering
\begin{tabular}{l c c c}
\hline\hline
\noalign{\smallskip}
Image & $\theta_1$ ($\arcsec$) & $\theta_2$ ($\arcsec$) & $\mu_{\rm macro}$ \\
\noalign{\smallskip}
\hline
\noalign{\smallskip}
A & $-1.188 \pm 0.001$           & $\phantom{-}1.321 \pm 0.001$ & $-10^{+1}_{-2}$ \\
B & $-1.805 \pm 0.002$           & $-0.802 \pm 0.002$           & $\phantom{-}6.3 \pm 0.3$ \\
C & $\phantom{-}0.210 \pm 0.003$ & $-1.606 \pm 0.003$           & $-3.2 \pm 0.2$ \\
D & $\phantom{-}2.738 \pm 0.004$ & $\phantom{-}1.046 \pm 0.004$ & $\phantom{-}3.35 \pm 0.08$ \\
E & $\phantom{-}0.680 \pm 0.010$ & $\phantom{-}2.530 \pm 0.010$ & $-1.6 \pm 0.3$ \\
\noalign{\smallskip}
\hline
\end{tabular}
\tablefoot{Positions are measured relative to the mean centre of light of G1, with $\theta_1$ oriented West and $\theta_2$ North.}
\end{table*}

\begin{figure}
    \centering
    \includegraphics[width=0.65\linewidth]{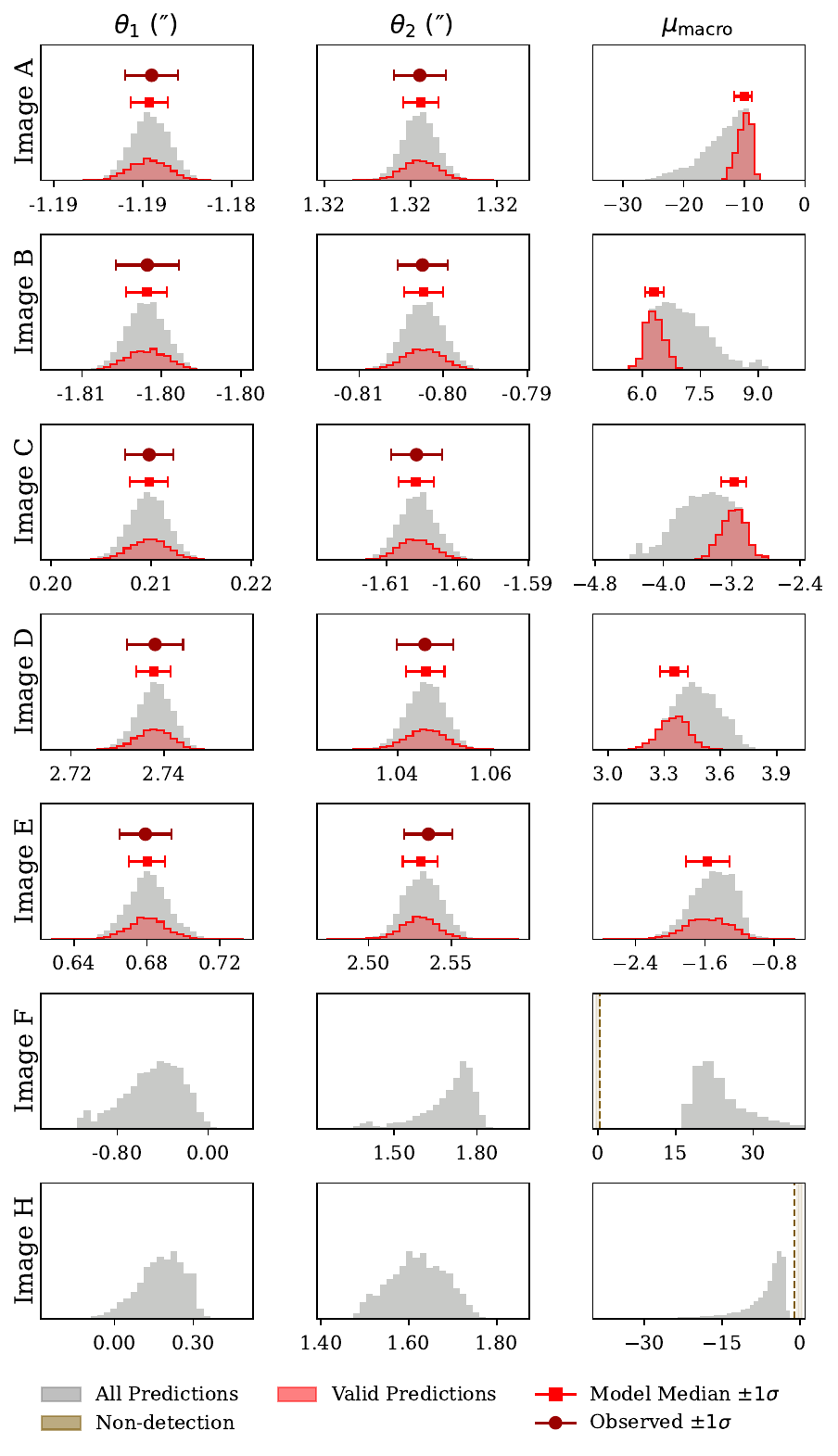}
    \caption{Posterior distributions of the predicted image positions ($\theta_{\rm 1, pred}$, $\theta_{\rm 2, pred}$) and macro-model magnifications ($\mu_{macro}$) for the single $K$ band \texttt{GLEE} model (Model~II). The grey histograms show the full distribution of all model predictions, using the five SN images as constraints without penalising models that predict additional images. The red histograms represent the subsets of models that validly reproduce the observed image multiplicity (=5) in the $K$ band. The lighter error bars with a square marker indicate the median and $1\sigma$ intervals for this subset, while the darker error bars with a circle marker show the observed astrometry with $1\sigma$ uncertainties. The bottom rows display the distributions for the additional counter-images F and H, which are predicted to be magnified ($|\mu_{\rm macro}| > 0.1$) but are not detected in the observations. The region $|\mu_{\rm macro}| \leq 0.1$, where an image may be present but obscured by noise, is highlighted in brown.}
    \label{fig:placeholder}
\end{figure}

\FloatBarrier 
\begin{figure*}[h!]
    \centering
    \includegraphics[width=\linewidth]{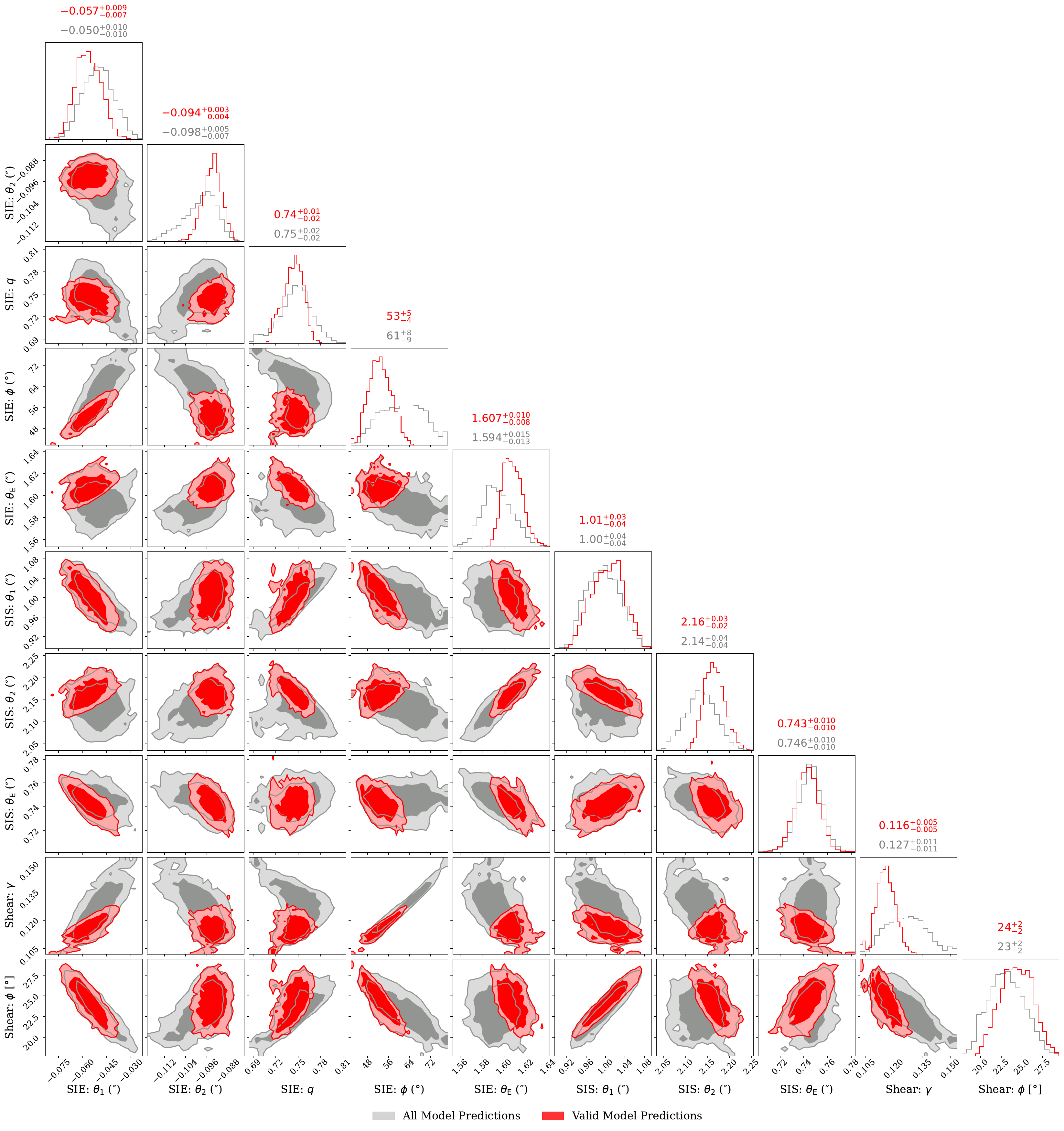}
    \caption{Posterior distributions of the lens model parameters for the single K-band GLEE model (Model~II). The grey histograms show the full distribution of all model predictions. The coloured histograms (red) represent the subset of models that validly reproduce the observed image multiplicity. The numerical values above the diagonal columns indicate the median and 1$\sigma$, while the off-diagonal panels display the 2D covariances between the model parameters.}\label{fig:glee_corner_sb}
\end{figure*}

\clearpage

\end{appendix}
\end{document}